\documentclass[reprint,aps,pra]{revtex4-1}
\usepackage{color,graphicx,calc,url}
\usepackage{dcolumn}
\usepackage{bm}
\usepackage{amssymb}
\usepackage{amsmath}
\usepackage{wasysym}
\usepackage{mathrsfs}

\begin{document}
\title{Tunnel magneto-Seebeck effect}
\author{T.~Kuschel$^1$, M. Czerner$^2$, J. Walowski$^3$, A.~Thomas$^4$, H.~W.~Schumacher$^5$, G.~Reiss$^1$, C.~Heiliger$^2$, M.~M\"unzenberg$^3$ \email{Electronic mail: tkuschel@physik.uni-bielefeld.de}}
\affiliation{$^1$Center for Spinelectronic Materials and Devices, Department of Physics, Bielefeld University, Universit\"atsstra\ss e 25, 33615 Bielefeld, Germany\\
$^2$ Institut f\"ur theoretische Physik, Justus Liebig University Giessen, Heinrich-Buff-Ring 16, 35392 Giessen, Germany\\
$^3$ Institut f\"ur Physik, Greifswald University, Felix-Hausdorff-Strasse 6, 17489 Greifswald, Germany\\
$^4$ Leibniz Institute for Solid State and Materials Research Dresden (IFW Dresden), Institute for Metallic Materials, Helmholtzstrasse 20, 01069 Dresden, Germany\\
$^5$ Physikalisch-Technische Bundesanstalt, Bundesallee 100, 38116 Braunschweig, Germany}

\date{\today}


\begin{abstract}
The interplay of charge, spin and heat transport is investigated in the fascinating research field of spin caloritronics, the marriage of spintronics and thermoelectrics. Here, many new spin-dependent thermal transport phenomena in magnetic nanostructures have been explored in the recent years. One of them is the tunnel magneto-Seebeck (TMS) effect in magnetic tunnel junctions (MTJs) that has large potential for future nanoelectronic devices, such as nanostructured sensors for three-dimentional thermal gradients, or scanning tunneling microscopes driven by temperature differences. The TMS describes the dependence of the MTJ's thermopower on its magnetic configuration when a thermal gradient is applied. In this review, we highlight the successful way from first observation of the TMS in 2011 to current ongoing developments in this research area. We emphasize on different heating techniques, material designs, applications, and additional physical aspects such as the role of the thermal conductivity of the barrier material. We further demonstrate the efficient interplay between ab initio calculations and experiments within this field, as this has led, e.g., to the detection of large TMS ratios in MTJs with half-metallic Heusler electrodes.
\end{abstract}

\maketitle

\section{General introduction}

\begin{figure*}[t!]
\centering
\includegraphics[width=1 \linewidth]{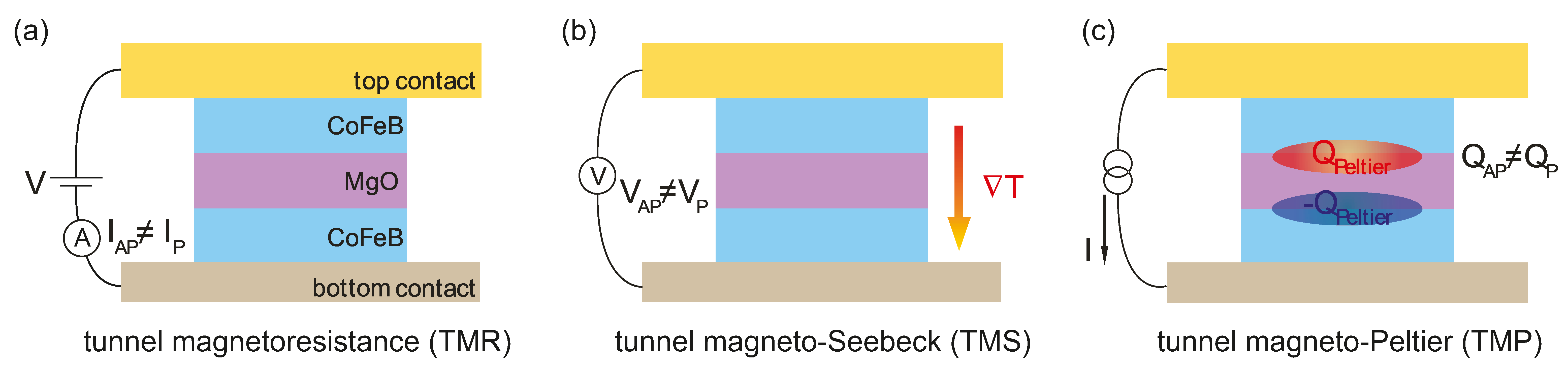}
\caption{MgO-based MTJ design with Co-Fe-B magnetic electrodes for the demonstration of (a)~tunnel magnetoresistance~(TMR) with applied voltage $V$ and resulting charge current $I_{\textrm{AP/P}}$ leading to the tunnel resistances $R_{\textrm{AP/P}}=V/I_{\textrm{AP/P}}$, (b)~tunnel magneto-Seebeck (TMS) effect with applied thermal gradient $\nabla T$ and resulting thermopower voltage $V_{\textrm{AP/P}}$ described by the Seebeck coefficients $S_{\textrm{AP/P}}=V_{\textrm{AP/P}}/\Delta T$ as well as (c)~tunnel magneto-Peltier (TMP) effect with applied charge current $I$ and resulting heat accumulation $Q_{\textrm{AP/P}}$ depending on the antiparallel (AP) or parallel (P) alignment of the Co-Fe-B electrodes. Schematic taken from Ref.~\cite{Shan2015} and modified.}
\label{fig:TMR-TMS-TMP}
\end{figure*}

In the emerging field of spin caloritronics~\cite{Bauer2012,Muenzenberg2012,Boona2014,Yu2017} a huge amount of new spintronic effects related to thermoelectrics has been explored. The heart of spin caloritronics is the generation of spin-dependent phenomena by thermal means. Analog to the classical Seebeck effect~\cite{Seebeck1822}, which is the generation of a thermoelectric voltage by the application of a temperature gradient, the spin-dependent Seebeck effect in magnetic metals and nanostructures is the thermovoltage (or thermocurrent) that depends on the magnetic state of the material or nanodevice. Here, the initial experiments~\cite{Gravier2006,Uchida2008,Slachter2010} that started the spin caloritronics research field have been triggered by early magnetothermopower experiments in multilayers~\cite{{Shi1993,Bailya2000}}. One key spin-dependent phenomenon is the tunnel magneto-Seebeck (TMS) effect that describes the induced magnetothermopower in a magnetic tunnel junction (MTJ) (see Fig.~\ref{fig:TMR-TMS-TMP}(b)). The TMS has been theoretically predicted by Czerner et al.~\cite{Czerner2011} and experimentally observed independently by Walter~\cite{Walter2011} et al. as well as by Liebing et al.~\cite{Liebing2011} for MgO-based MTJs in 2011. The first large effect in AlOx-based MTJs has been reported by Lin et al.~\cite{Lin2012}.

This fascinating spin-dependent phenomenon is the spin caloritronic analog to the tunnel magnetoresistance (TMR) that has now been explored for decades in spintronics~\cite{Julliere1975} (see Fig.~\ref{fig:TMR-TMS-TMP}(a)). Although both TMR and TMS have their origin in the spin-dependent transmission function of the MTJ and, therefore, are related to the spin-split density of states (DOS) of the involved materials, the magnitudes of both effects have no direct connection. While the TMR scales with the imbalance of majority and minority spin states close to the Fermi level, the TMS is affected by the spin-dependent asymmetry of the DOS with respect to the Fermi energy. This theoretical description has recently been proven experimentally by detecting a large TMS in MTJs with half-metallic Heusler alloys that have the Fermi energy close to one edge of the half-metallic band gap, thus supporting the TMS by a large asymmetry of the DOS with respect to the Fermi energy~\cite{Boehnke2017}.

The inverse effect of the TMS, the tunnel magneto-Peltier (TMP) effect has been discovered already. It describes the generation of a spin-dependent temperature gradient across an MTJ when a charge current is applied~\cite{Shan2015} (see Fig.~\ref{fig:TMR-TMS-TMP}(c)). This effect together with the TMS effect fulfills the Onsager reciprocity~\cite{Onsager1931} for spin-heat transport conversion. Parallel to the first TMS experiments in 2011, thermal spin injection into semiconductors has been discovered also known as Seebeck spin tunneling~\cite{LeBreton2011,Jansen2012,Jeon2014,Jeon2015}.

The TMS in combination with a spin-transfer torque (STT)~\cite{Slonczewski1989} has the potential to thermally assist the magnetic switching of MTJs~\cite{Pushp2015}. After the first theoretical descriptions of the thermal STT in spin valves~\cite{Hatami2007,Slonczewski2010} and MTJs~\cite{Jia2011,Heiliger2014} first evidences towards an experimental confirmation can be found in literature~\cite{Yu2010,Leutenantsmeyer2013,Zhang2016,Liebing2016,Cansever2018}. While the electrically induced STT (magnetic switching by spin-polarized charge currents) is already utilized in commercially available STT-based magnetoresistive random-access memory devices, the thermal STT is not yet explored in a way that it can be used for commercial devices. Different applications are currently more promising such as three-dimensional sensing of thermal gradients~\cite{Martens2018} or scanning tunneling microscopy with temperature differences~\cite{Friesen2018a,Friesen2018b} as presented in the following.

Within this review we want to illustrate the TMS from first observation over improvement of devices and TMS magnitudes to future applications. In addition, we discuss fundamental open questions and ongoing research such as the accurate determination of the thermal gradient in the MTJs and the role of the thermal conductivity of the insulating barrier. This review starts with an introduction into the topic of TMR and TMS (Sec.~\ref{intro}) including the description of typical TMS experiments subdivided by the individual heating technique (laser-, electric-, intrinsic heating). Section~\ref{improvements} demonstrates that applying a bias voltage in addition to the temperature gradient can enhance the TMS up to 3000\%. It further discusses different electrode and barrier materials as for example MgAl$_2$O$_4$ to optimize the TMS output. The final Sec.~\ref{aspects} presents further aspects of the TMS, such as its angular dependence, the determination of the thermal conductivity of the tunnel barrier material and additional (magneto)thermoelectric effects that have to be taken into account. We finally discuss in that section the influence of three-dimensional thermal gradients and the use of temperature differences in scanning tunneling microscopy.

\section{Introduction to TMR and TMS}
\label{intro}

MTJs have always been important building blocks in spin- or magneto-electronics~\cite{Prinz1998}. The interest started with the discovery of TMR at room temperature in 1995~\cite{Moodera1995,Miyazaki1995}. Now, MTJs are used in magnetic logic devices, magnetic sensors and magnetoresistive random-access memories~\cite{Prinz1998,Miao2011}. The prediction~\cite{Mathon2001,Butler2001} and experimental realization~\cite{Parkin2004,Yuasa2004} of giant TMR in MgO-based MTJs as well as current induced magnetic switching (caused by STT) in 2004~\cite{Huai2004} further stimulated worldwide research activities. 

An MTJ consists of a trilayer, in which two ferromagnetic electrodes are separated by a thin insulating tunnel barrier. A bias voltage applied across the tunnel barrier results in a current perpendicular to the layer plane~\cite{Moodera1999}. A schematic of an MTJ is given in Fig.~\ref{fig:TMR-TMS-TMP} for the individual spin-dependent effects mentioned so far. 

The two extreme configurations of the electrodes' magnetization are parallel (P) and antiparallel (AP) alignment. In most cases, the tunnel current is smaller in the AP- than in the P-case related (positive TMR). However, for some materials the opposite case can occur connected to a negative TMR~\cite{Sharma1999,DeTeresa1999,Klewe2013,Marnitz2015}. Generally, the TMR ratio given by the difference of the two conductivities $g_\textrm{P}$ and $g_\textrm{AP}$ is
\begin{equation}
\textrm{TMR} = \frac{g_{\textrm{P}}-g_{\textrm{AP}}}{\min (|g_{\textrm{P}}|,|g_{\textrm{AP}}|)}\ \ \ .
\label{eq:def:TMR}
\end{equation}

So far, the TMR ratio reaches values of up to 604\% in MgO-based MTJs at room temperature~\cite{Ikeda2008}. Typical TMR magnetic field loop measurements for Co-Fe-B/MgO/Co-Fe-B MTJs are presented in Figs.~\ref{fig:TMSloops}(a) and (c) taken from the original TMS publications of Walter et al.~\cite{Walter2011} and Liebing et al.~\cite{Liebing2011}. While the TMR curve of Fig.~\ref{fig:TMSloops}(a) describes a major loop (magnetization of both electrodes switch successively), the TMR curve of Fig.~\ref{fig:TMSloops}(c) represents a minor loop (magnetization of just one electrode switches) due to exchange-biased magnetic pinning of one electrode.

The size of the TMR is connected to the spin polarization of the two electrodes and can be described by the model of Julli$\acute{\textrm{e}}$re~\cite{Julliere1975}. The model uses Fermi's golden rule to calculate the TMR from the spin polarizations $P_1$ and $P_2$ of the two electrodes. It concludes
\begin{equation}
\text{TMR} = \frac{2P_1P_2}{1+P_1P_2}\ \ \ .
\end{equation}

The model is useful for a first understanding of the TMR and can give some hints for the experimentalist to design suitable materials. For example a half-metal, a material with only one kind of spin polarization at the Fermi level, is expected to lead to very large TMR ratios. However, the model assumes constant transmission probability for all electrons, which is not generally the case. For example for MgO-based MTJs, the situation becomes more elaborate, because of the complex band structure of the magnesia in contact with the two electrodes. Now, wave functions of distinct symmetry decay further into the barrier than others, leading to different transmission probabilities for electronic wave functions with varying symmetry~\cite{Butler2001}. This violates one of the main assumptions of the Julli$\acute{\textrm{e}}$re-model, which therefore cannot be used in case of MTJs with symmetry filtering properties.

If we exchange the applied bias voltage by a temperature gradient and replace the measured current by the thermovoltage as theoretically suggested in early literature~\cite{Czerner2011,Wang2001,McCann2002a,McCann2002b}, we can observe the TMS~\cite{Walter2011,Liebing2011} as displayed in Fig.~\ref{fig:TMR-TMS-TMP}(b). The TMS depends on the magnetic configuration of the MTJ as the TMR does. However, the full magnetic field angular dependence is different as we will discuss in Sec.~\ref{sec:angular-dep}. Analog to the TMR, we define the TMS as 
\begin{equation}
\textrm{TMS} = \frac{S_{\textrm{P}}-S_{\textrm{AP}}}{\min (|S_{\textrm{P}}|,|S_{\textrm{AP}}|)} \ \ \ .
\label{eq:def:TMS}
\end{equation}

Here, $S_{\textrm{P}}$ and $S_{\textrm{AP}}$ are the Seebeck coefficients for P and AP magnetic orientation of the MTJ. In contrast to the electric conductivity, the Seebeck coefficients can generally change sign. Hence, the TMS can have divergences whenever one of the Seebeck coefficients is zero~\cite{Czerner2011}. 

If the transmission functions $T_{\textrm{P}}(E)$ and $T_{\textrm{AP}}(E)$ for the two magnetic configurations and the occupation function $f(E,\mu,T)$ are known, the Seebeck coefficients $S_{\textrm{P}}$ and $S_{\textrm{AP}}$ can be calculated via the moments $L_{\textrm{n}}$ obtained from the Landauer formalism~\cite{Czerner2011,Ouyang2009}. These moments are defined as
\begin{equation}
  L_{\textrm{n}}=\frac{2}{h} \int T(E) (E-\mu)^n (-\partial_{\textrm{E}} f(E,\mu,T)) dE \ \ \ ,
  \label{eq:Ln}
\end{equation}
which are different for the P and AP case. The conductances $g_{\textrm{P}}$ and $g_{\textrm{AP}}$ as well as the Seebeck coefficients $S_{\textrm{P}}$ and $S_{\textrm{AP}}$ can be derived as
\begin{equation}
\begin{aligned}
   g=e^2 \: L_0 \ \ \ \textrm{and}\ \ \ S=-\frac{1}{e T} \frac{L_1}{L_0} \ \ \ .
  \end{aligned} 
	\label{eq:g-and-S}
\end{equation}

From these equations one can conclude that the conductance $g$ is proportional to the integral of the function 
\begin{equation}
T(E) \cdot \left ( - \frac{\partial f(E)}{\partial E} \right ) \ \ \ ,
\label{eq:integrand}
\end{equation} 
whereas the Seebeck coefficient $S$ is proportional to the center of mass of this function. Since $g$ and $S$ are not directly related, TMR and TMS do not have a direct connection. However, they are determined by the same $T(E)$ and, thus, are based on the same spin-split DOS of the materials involved. 

For the theoretical approach, ab initio calculations based on density-functional theory (DFT) have been conducted. In particular, the Korringa-Kohn-Rostoker (KKR) and the non-equilibrium Green's function method has been used to obtain the transmission function $T(E)$~\cite{Czerner2011} that leads to the conductance and Seebeck coefficients via Eqs.~(\ref{eq:Ln}) and (\ref{eq:g-and-S}) as well as to the TMR and TMS values via Eqs.~(\ref{eq:def:TMR}) and (\ref{eq:def:TMS}). For modeling the electrode material at the interface to the tunnel barrier, a supercell approach has been applied in the beginning~\cite{Czerner2011,Walter2011,Czerner2012}. In subsequent studies the coherent potential approximation has been employed in the KKR method~\cite{Heiliger2013a}, which has less computational effort and the possibility to use an arbitrary composition of the leads in contrast to the supercell approach. It turned out that non-equilibrium vertex corrections have to be taken into account within the KKR code~\cite{Franz2013}. Further theoretical descriptions and aspects can be found in the mentioned references summarized in a recent review about spin caloric transport from DFT by Popescu et al.~\cite{Popescu2018}.

In the following we will discuss the TMS for different kind of heating techniques. The initial TMS observation in MTJs has been achieved simultaneously in two groups. While Walter et al.~\cite{Walter2011} used a laser to heat the top part of the MTJ, Liebing et al.~\cite{Liebing2011} patterned nanosized heater wires on top of the MTJ. Later on, Zhang and Teixeira et al.~\cite{Zhang2012a,Teixeira2013} presented an intrinsic heating approach using the applied tunnel current itself that creates a thermal gradient and thus an intrinsic TMS. This latter approach was discussed by Huebner et al.~\cite{Huebner2016} and compared to laser-heating induced TMS as recapped in Sec.~\ref{sec:intrinsic-TMS} of this review.

\subsection{Laser-heating induced TMS}
\label{sec:laser-heat-TMS}
In the first laser-heating induced TMS experiments, a $784\,\textrm{nm}$ laser has been used with a focus of $15\,\mu\textrm{m}$ to $20\,\mu\textrm{m}$ and a power of up to $100\,\textrm{mW}$ to heat up the top part of the MTJ and, thus, to generate the temperature gradient~\cite{Walter2011}. In later experiments different laser wavelengths (e.g. $638\,\textrm{nm}$), smaller focus down to $2\,\mu\textrm{m}$ and larger powers of up to $150\,\textrm{mW}$ have been employed~\cite{Boehnke2017,Boehnke2013,Boehnke2015,Martens2017}. The Co-Fe-B/MgO/Co-Fe-B MTJs with a size of $1\,\mu\textrm{m}$ to $12.5\,\mu\textrm{m}$ have been patterned by electron beam lithography and ion beam etching as well as post-annealing~\cite{Walter2011}.

\begin{figure}[t!]
\centering
\includegraphics[width=1 \linewidth]{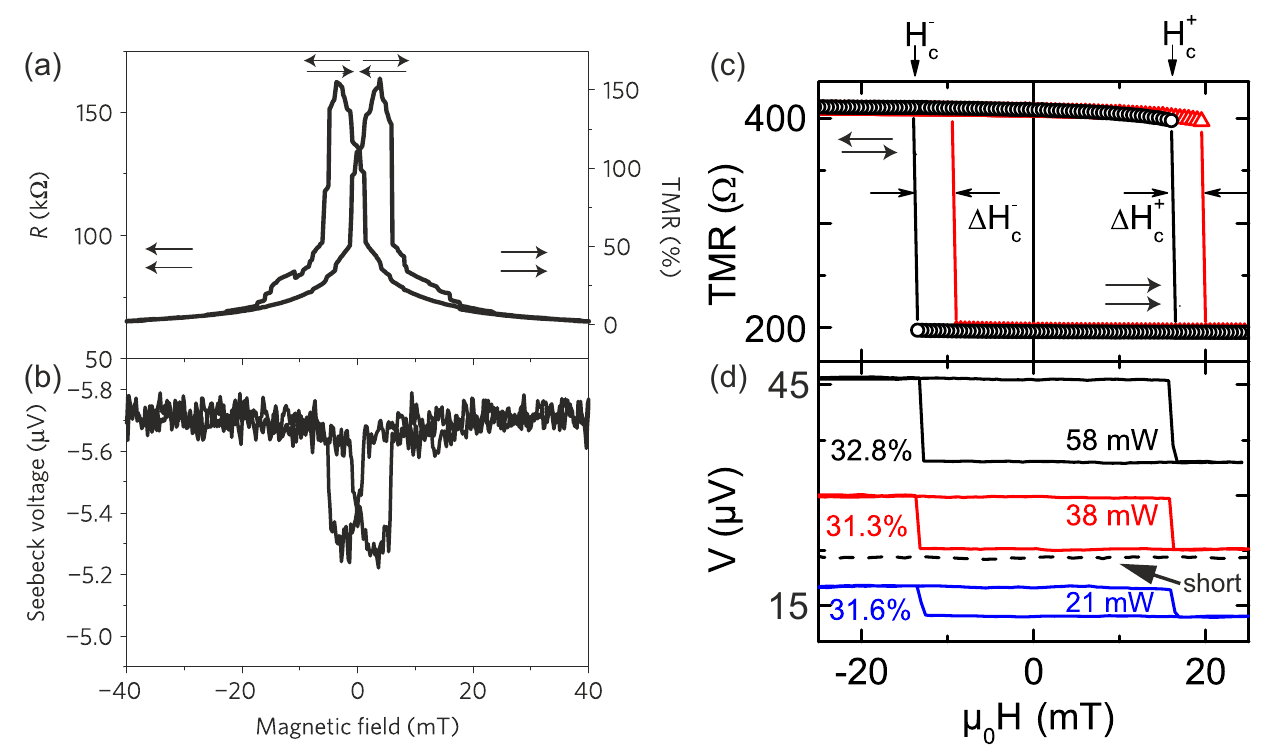}
\caption{TMR and TMS voltage vs. magnetic field loops of Co-Fe-B/MgO/Co-Fe-B MTJs. (a) TMR major loop with different resistances for the P and AP configuration of the MTJ. (b) Laser-heating induced TMS major loop. Figure taken from Ref. \cite{Walter2011}. (c) TMR minor loop. The magnetic field $H_{\textrm{HL}}$ from the heater line results in a shift $\Delta H_{\textrm{C}}^{\pm}$ of the coercive field. (d) TMS minor loop for different heating powers of the heater line. The dashed line indicates the residual magnetic-field-independent voltage after the dielectric breakdown of the MTJ. Figure taken from Ref.~\cite{Liebing2011}.}
\label{fig:TMSloops}
\end{figure}

A typical TMR and TMS result can be found in Figs.~\ref{fig:TMSloops}(a) and (b), respectively. The arrows indicate the magnetic alignment of the electrodes of the MTJ. The TMR curve shows a hard-soft switching with a positive TMR and 150\% effect, while the TMS for a laser power of $30\,\textrm{mW}$ loop has a drop in Seebeck voltage of $-8.8\%$ going from $5.7\,\textrm{mV}$ (P) to $5.3\,\textrm{mV}$ (AP) thermovoltage.

In order to estimate the temperature difference across the barrier, finite element modeling based on the experimental parameters (laser focus and power, material properties such as the thermal conductivity) has been done as described in Ref.~\cite{Walter2011}. Thermal interface resistances and related effects have been neglected so far, but will be discussed in Sec.~\ref{sec:thermal-resistivity} of this review. A temperature difference of $\Delta T=53\,\textrm{mK}$ has been derived for a laser focus of $15\,\mu\textrm{m}$, a laser power of $30\,\textrm{mW}$ and an MgO barrier thickness of $2.1\,\textrm{nm}$ using the thin film value of $\kappa=4\,\textrm{W/mK}$ for the thermal conductivity of the MgO barrier~\cite{Walter2011}. The resulting $S_{\textrm{P}}=V_{\textrm{P}}/\Delta T=-107.9\,\mu\textrm{V/K}$ and $S_{\textrm{AP}}=V_{\textrm{AP}}/\Delta T=-99.2\,\mu\textrm{V/K}$ lead to the TMS of $-8.8\%$ following Eq.~(\ref{eq:def:TMS}). 

\begin{figure}[t!]
\centering
\includegraphics[width=1 \linewidth]{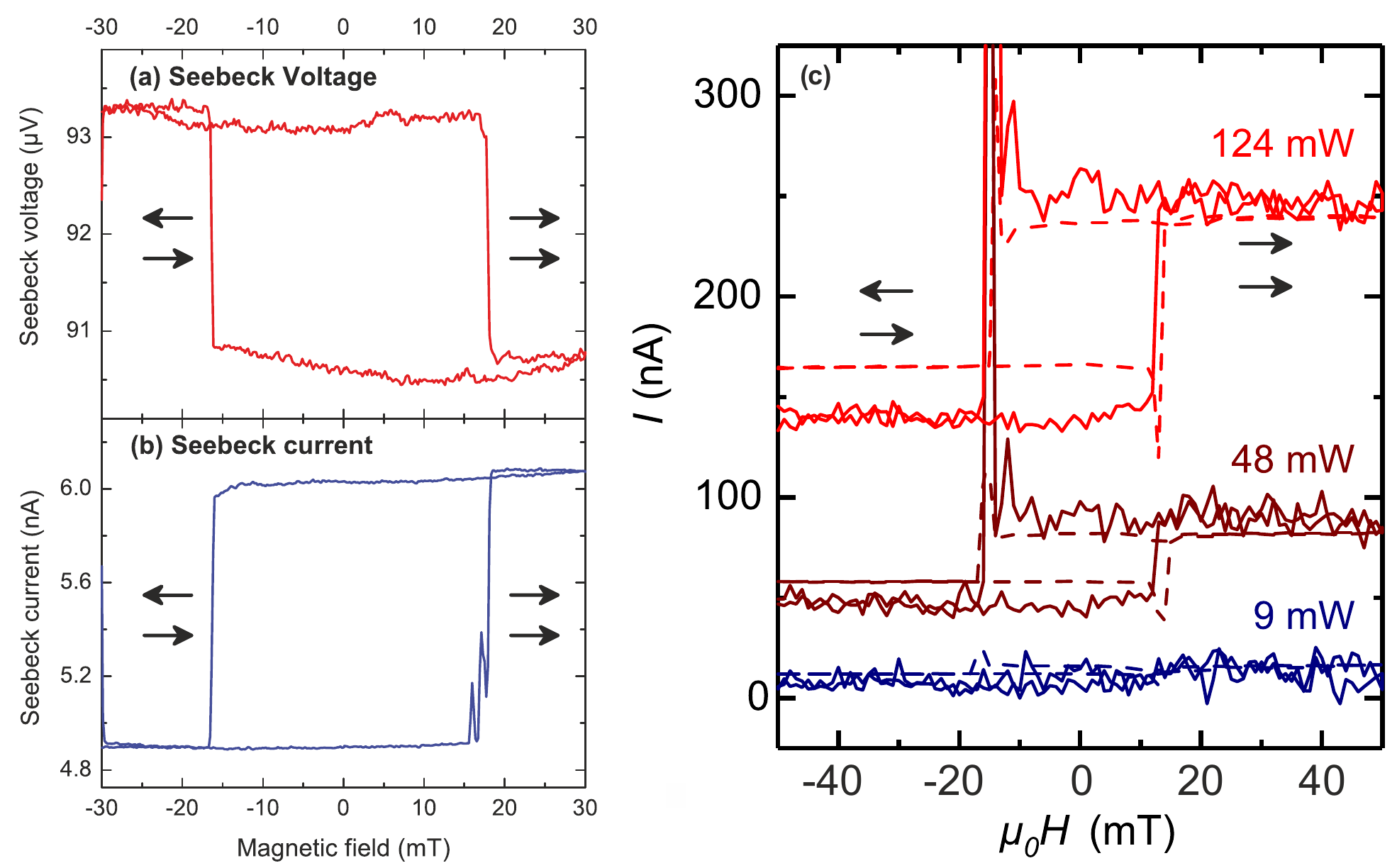}
\caption{TMS current vs. magnetic field loops of Co-Fe-B/MgO/Co-Fe-B MTJs. (a) Laser-heating induced TMS voltage. (b) Laser-heating induced TMS current. Figure taken from Ref.~\cite{Boehnke2013}. (c) TMS current loop for different heating powers of an external nanostructured heater. Figure taken from Ref.~\cite{Liebing2012}.}
\label{fig:TMSloopscurrent}
\end{figure}

Another way to determine the thermoelectric properties of these devices is to measure the thermoelectrically generated current of the device by a current amplifier in closed circuit conditions. A typical measurement of the laser-heating induced TMS current is shown in Fig.~\ref{fig:TMSloopscurrent}(b) and can be compared to the laser-heating induced TMS voltage via the known tunnel resistance (Fig.~\ref{fig:TMSloopscurrent}(a)). Note that the TMS shows an increase of the Seebeck current in the parallel state opposite to the Seebeck voltage behavior.

In general, the position and size of the laser strongly affects the TMS signal~\cite{Boehnke2013,Martens2017}. It was found that the TMS vanishes for laser positions far off the centered position on the MTJ and maximizes if aligned at the center. For slight off-positions at the edges of the MTJs, additional heat transport effects such as the anomalous Nernst effect arise~\cite{Martens2018}. They will be discussed in Sec.~\ref{sec:ANE} of this review. Any temperature dependence of the TMS has only been rarely studied so far~\cite{Walter2011}. Theoretical calculations have been carried out and discuss sign changes of the TMS at specific temperatures depending on the Co-Fe interface termination~\cite{Czerner2012} and composition~\cite{Heiliger2013a}. Still systematic experimental temperature-dependent studies have to be done to confirm the theoretical predictions.

In the beginning, only MTJs with MgO tunnel barrier have been explored. In 2012, Lin et al. reported on laser-heating induced TMS studies in MTJs with aluminium oxide barrier~\cite{Lin2012} with similar values for TMR and TMS of about 40\%. The TMS is mostly independent from the heating power comparable to the results for MgO-barrier MTJs~\cite{Liebing2011}. Furthermore, Lin et al.~\cite{Lin2012} were able to invert the thermal gradient by heating either the top or bottom contact of the MTJ, thus inverting the detected thermocurrent. However, they could not observe any magnetic field dependence for the thermocurrent. This can be expected if the larger thermovoltage in the AP case is compensated by the simultaneously increased tunnel resistance. Later on, further theoretical studies on MTJs with aluminium oxide barrier have been conducted by L\'{o}pez-Mon\'{i}s et al.~\cite{Lopez-Monis2014}. Huebner et al. introduced MgAl$_2$O$_4$ tunnel barriers for laser-heating induced TMS investigations~\cite{Huebner2016,Huebner2017,Huebner2018}. This will be further discussed in the Secs.~\ref{sec:intrinsic-TMS} and~\ref{sec:variation-material} of this review.

\begin{figure}[t!]
\centering
\includegraphics[width=1 \linewidth]{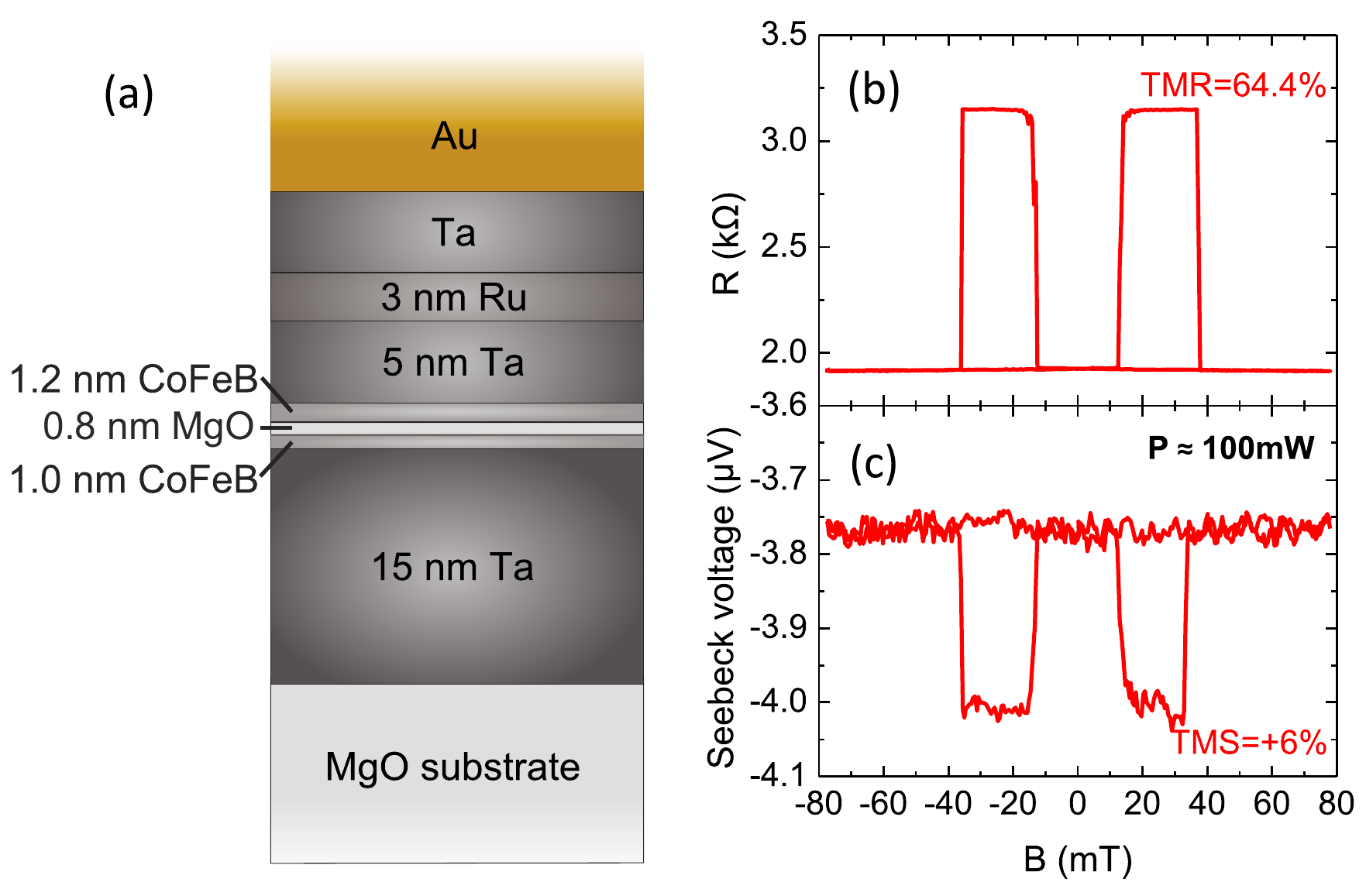}
\caption{Laser-heating induced TMS voltage for an MTJ with PMA. (a) Schematic of the MTJ's structure. (b) TMR magnetic field loop. (c) TMS voltage loop for a laser power of $100\,\textrm{mW}$.}
\label{fig:PMA-TMS}
\end{figure}

The most laser-heating induced TMS studies so far concentrate on MTJs with in-plane magnetized electrodes. However, some rare experiments have been done with MTJs exhibiting a perpendicular magnetic anisotropy (PMA). This type of MTJs is also interesting for STT switching experiments, since it usually needs less current density to manipulate the magnetization of one electrode by the spin-polarized current of the other. Thus, MTJs with PMA are also promising candidates for thermal STT experiments~\cite{Leutenantsmeyer2013}. A typical MTJ stack with very thin barrier and electrode thicknesses supporting PMA is shown in Fig.~\ref{fig:PMA-TMS}(a). The resulting TMR ($64.4\,\%$) and TMS ($6\%$) curves are presented in Figs.~\ref{fig:PMA-TMS}(b) and (c), respectively. This kind of sample has been studied in the work of Leutenantsmeyer et al.~\cite{Leutenantsmeyer2013}.

\subsection{Electric-heating induced TMS}
\label{sec:elec-heat-TMS}
Another way to generate thermal gradients across the layer stack of an MTJ is the use of electric Joule heater lines (HL) lithographically placed on top of the MTJs. Electric heater schemes are well established to characterize in-plane thermoelectric properties of thin films~\cite{Mavrokefalos2007}. Such electric heaters on membrane based thermoelectric measurement platforms were use, e.g., to characterize the thermoelectric properties of magnetic thin films~\cite{Avery2012}. With respect to spin-dependent perpendicular transport electric heating was employed to characterize Seebeck spin tunneling from a ferromagnet into a semiconductor~\cite{LeBreton2011,Jeon2014} and spin heat accumulation in nanopillar spin valves~\cite{Dejene2013}. Such electric heater schemes have also been realized by Liebing et al. to study the TMS by detecting the tunnel magnetothermopower~\cite{Liebing2011} and tunnel magnetothermocurrent~\cite{Liebing2013} of Co-Fe-B/MgO/Co-Fe-B MTJ nanopillars.

In the first electric-heater induced TMS studies with in-plane magnetization, the MTJ stacks comprised a complex stack sequence including an antiferromagnetic pinning layer and a compensated synthetic antiferromagnetic reference layer~\cite{Serrano-Guisan2008}. Nanopatterning of the MTJ cells down to about $160\,\textrm{nm}$ as well as contact definition was realized by electron beam lithography and clean room processing. A $5\,\mu\textrm{m}$ wide HL was positioned on top of the nanopillar and separated from the MTJ top contact by a 160 nm thick Ta$_2$O$_5$ dielectric. Thermal gradients across the MTJ were generated by applying AC or DC heater currents of up to $I_{\textrm{heat}} = 60\,\textrm{mA}$ through the HL. Note that for electric heating the typical HL Oersted field of the order of 0.1 mT/mA must be considered for data analysis. The heater temperature during the experiments can be determined by calibrating the temperature dependent resistance change on a variable temperature probe station. This heater temperature and the temperature of the sample stage as input for finite-element modeling of the heat flux trough the complex layer stack can be utilized to estimate the temperature drop across the MgO barrier~\cite{Liebing2012}. In the above experiments, maximum temperature drops of $\Delta T\approx 45\,\textrm{mK}$ were obtained.

Figure~\ref{fig:TMSloops}(d) shows typical thermopower data of these MTJ nanopillars. The measured thermopower voltage $V$ in open circuit conditions between top and bottom contact of the MTJ is displayed for heater powers $P_{\textrm{heat}}$ of $21$, $38$ and $58\,\textrm{mW}$ as function of the magnetic easy axis field $H$. For all three curves $V_{\textrm{P}}$ in the P state is lower than $V_{\textrm{AP}}$ in the AP state with a maximum difference of $\Delta V\approx 11\,\mu\textrm{V}$. One can identify a TMS value independent from the heating power with an average $\frac{\Delta V}{V_{\textrm{P}}}\approx 32\%$. The black dashed line in Fig.~\ref{fig:TMSloops}(d) indicates the thermopower of the same MTJ patterned into nanopillars with diameter down to $160\,\textrm{nm}$ after the MgO barrier was applied to current stress (dielectric breakdown). The magnetic-field-independent $V_{\textrm{short}}$ allows to determine the thermopower contribution of all non-magnetic layers of the devices. Subtracting this background yields the true TMS value of the Co-Fe-B/MgO/Co-Fe-B MTJ reaching up to $90\%$ for the given samples.

The different TMS magnitudes and the opposite signs between laser- (Walter et al.~\cite{Walter2011}) and electric-heating (Liebing et al.~\cite{Liebing2011}) induced TMS detection in MgO-based in-plane magnetized MTJs can be explained by slight differences in the Co-Fe interface termination~\cite{Czerner2012} and composition~\cite{Heiliger2013a} of the MTJs. The theoretical work of Czerner and Heiliger shows that the influence of these electrode properties can have a huge impact on the TMS magnitude and sign. However, systematic experimental investigations varying only the interface termination or material composition while keeping the residual parameters of the MTJs constant is very challenging and could not be realized experimentally so far.

A typical example of the electric-heating induced TMS current is shown in Fig.~\ref{fig:TMSloopscurrent}(c). The experimental data can be well modeled taking into account the Onsager transport equations and the TMR of the MTJ devices~\cite{Johnson1987,Johnson2010}. The electric current $I$ in the presence of the voltage $V$ and a temperature gradient $\nabla T$ across the MTJ is given by $I=g\,V-g\,S\,\nabla T$ where $g$ is again the electric conductance of the MTJ and $S$ the Seebeck coefficient. Measurements of the junction resistance $R$ as well as $V$ and $I$ in an open circuit ($I=0$) and closed circuit ($V=0$) configuration, respectively, allow to determine $g=R^{-1}$ and $V=S\,\nabla T$. Using these values, the predicted magnetic thermocurrent $I$ in closed circuit configuration is given by $I=\frac{V}{R}$ and shows a good agreement with the experimental data (solid lines in Fig.~\ref{fig:TMSloopscurrent}(c)).

B\"ohnert et al. extended these studies by the introduction of integrated thermometers allowing a better determination of the thermal gradient~\cite{Boehnert2017b} and a more detailed analysis of the thermal interface transport (see Sec.~\ref{sec:thermal-resistivity}). They further varied the MgO thickness~\cite{Boehnert2017a,Boehnert2018} as summarized in Sec.~\ref{sec:variation-barrier} of this review. So far, no electric-heater induced TMS experiments on MTJs with PMA have been conducted.

\subsection{Intrinsic TMS}
\label{sec:intrinsic-TMS}
Beside laser- and electric-heating another method for the generation of a temperature gradient across an MTJ has been suggested by Zhang and Teixeira et al.~\cite{Zhang2012a,Teixeira2013}, namely an intrinsic heating by the tunnel current itself leading to an intrinsic TMS. In their articles, they explain that the tunnel current $I$ induces the thermal gradient via Joule heating and describe the resulting intrinsic thermopower effect as a nonlinear correction to Ohm's law. They write 
\begin{equation}
  V(I)=R\,I+S\,\alpha\,I^2
  \label{eq:nl-Ohm}
\end{equation}

with the resistance $R$ and the Seebeck coefficient $S$ of the MTJ as well as $\alpha=\sum_{\textrm{j}}\,\eta_{\textrm{j}}\,R_{\textrm{j}}\,R_{\kappa_{\textrm{j}}}$. Here, $\eta_{\textrm{j}}$ is the thermal asymmetric parameter~\cite{Zhang2012a}, $R_{\textrm{j}}$ the resistance and $R_{\kappa_{\textrm{j}}}$ the thermal resistance of the j-th layer. Thus, it is assumed that any quadratic-in-$I$ contribution in the V-I curve of the MTJ can be related to this higher order term induced by a heating effect that modifies the measured voltage by an intrinsic TMS. However, this approach neglects any nonlinearity from the MTJ characteristics itself, e.g. barrier potential asymmetries.

In order to extract the intrinsic magneto-Seebeck coefficients from the V-I curve for P and AP alignment, the data is separated into odd ($(V_+-V_-)/2\propto I$) and even ($(V_++V_-)/2\propto I^2$) parts plotted against $I$ and $I^2$, respectively. The slope of the linear regressions (see Eq.~(\ref{eq:nl-Ohm})) gives the intrinsic magneto-Seebeck coefficients if $\alpha$ is estimated by a numerical calculation of the thermal profile of the MTJ~\cite{Zhang2012a,Teixeira2013,Huebner2016}. With this approach Zhang and Teixeira et al. concluded to have TMS values of more than 1000\%. Ning et al.~\cite{Ning2017} adopted this procedure and concluded to have similar large intrinsic TMS magnitudes for MTJs with PMA.

In order to compare the intrinsic TMS obtained by this symmetry analysis to the laser-heating induced TMS, Huebner et al.~\cite{Huebner2016} followed the same procedure for MgO and MgAl$_2$O$_4$ (MAO) based MTJs. It turned out that in their case, the separation of odd and even parts of the V-I curves does not necessarily give linear dependencies on $I$ and $I^2$, respectively. Therefore, a linear regression could only be made for small tunnel currents. The Seebeck coefficients obtained for the intrinsic TMS differ from the laser-heating induced Seebeck coefficients by more than one order of magnitude and have different sign. As an example for MgO-MTJs, they determined $S_{\textrm{P}}=0.3\,\mu\textrm{V/K}^{-1}$ and $S_{\textrm{AP}}=-7.5\,\mu\textrm{V/K}^{-1}$ for the intrinsic TMS via the symmetry analysis, while the laser-heating induced TMS results in $S_{\textrm{P}}=-1010\,\mu\textrm{V/K}^{-1}$ and $S_{\textrm{AP}}=-1320\,\mu\textrm{V/K}^{-1}$. Beside quite different Seebeck coefficients, the intrinsic TMS values of 105\% and -75\% for MgO and MAO based MTJs vary, as well, compared to the laser-heating induced TMS ratios of 23\% and 3.3\% measured on the same MTJs, respectively.

Alternatively, the higher orders in the V-I curve can be described by an asymmetric barrier potential without any connection to thermoelectrics. Using the Brinkman model~\cite{Brinkman1970}, one can fit the curves varying the potential height $\varphi$, asymmetry $\Delta \varphi$ and thickness $d$ of the barrier. The model is valid for MTJs with low or vanishing symmetry filter effect (constant transmission probability for all electrons). As discussed by Huebner et al., for this kind of MTJs the Brinkman model alternatively explains the higher orders in the V-I curves with reasonable parameters~\cite{Huebner2016}. They conclude, that the intrinsic TMS cannot be unambiguously identified by the symmetry analysis proposed by Zhang, Teixeira et al., because of the higher-order terms in the V-I curve that are already present due to the barrier potential of the MTJ itself. Additional techniques and analyses have to be developed in the future to clearly distinguish between V-I characteristics from any intrinsic TMS contribution and from the barrier potential of the MTJ.

\section{Increase of TMS ratio and thermopower}
\label{improvements}
There have been several ideas to increase the TMS magnitude. This includes the theoretical variation and optimization of the interface termination~\cite{Czerner2012} and alloy composition~\cite{Heiliger2013a} in case of Fe-Co electrodes. The latter basically shifts the Fermi level. Therefore, an idea from experiment came up that the same goal could be reached by applying a bias voltage~\cite{Boehnke2015}. This bias-enhanced TMS has been studied theoretically and experimentally and will be discussed next, followed by further material optimization aspects such as the use of half-metal Heusler compounds as electrodes~\cite{Boehnke2015} or varying the thickness and material of the tunnel barrier~\cite{Huebner2017,Boehnert2017a,Boehnert2018}.

\subsection{Bias-enhanced TMS}
\begin{figure}[b!]
\centering
	\includegraphics[width=1 \linewidth]{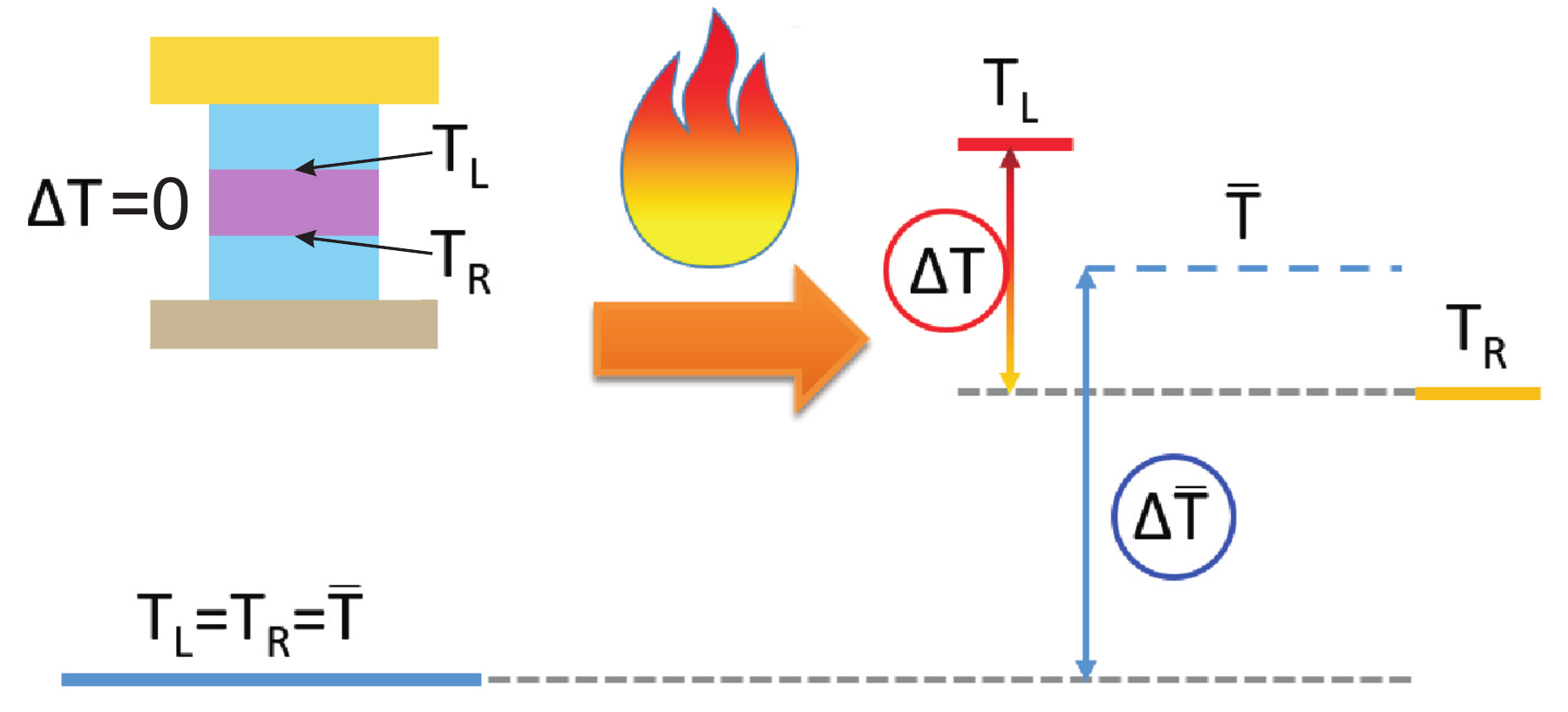}
	\caption{Temperature changes during heating the MTJ. A temperature gradient with the difference $\Delta T=T_{\textrm{L}}-T_{\textrm{R}}$ is building up and the mean temperature $\bar{T}=(T_{\textrm{L}}+T_{\textrm{R}})/2$ is increased by $\Delta \bar{T}$.}
	\label{fig:bTMS:temp}
\end{figure}

The idea of shifting the Fermi level by a bias voltage is triggered by typical STM experiments in where the electronic states can be selected by a bias voltage. In the case of the TMS it is more complicated, because there is not only the bias voltage but also the temperature gradient. Further, because now a bias voltage is applied the thermopower voltage due to the temperature gradient cannot be measured directly. To overcome this issue the thermocurrent rather than the thermovoltage is detected as discussed in Secs.~\ref{sec:laser-heat-TMS} and~\ref{sec:elec-heat-TMS} of this review, but with and without temperature gradient. Applying a temperature gradient not only leads to a temperature drop across the MTJ but also to an overall increase of average temperature. This is illustrated in Fig.~\ref{fig:bTMS:temp}. 

Thereby, the mean temperature $\bar{T}=(T_{\textrm{L}}+T_{\textrm{R}})/2$ changes by $\Delta \bar{T}$ whereas the temperature difference across the barrier is $\Delta T=T_{\textrm{L}}-T_{\textrm{R}}$ for $T_{\textrm{L}}$ and $T_{\textrm{R}}$ being the temperature at the upper and lower interface of the barrier, respectively (cf. Fig.~\ref{fig:bTMS:temp}). As shown in the examples of Secs.~\ref{sec:laser-heat-TMS} and \ref{sec:elec-heat-TMS}, simulations with finite element methods show that $\Delta T$ is in the order of 50 mK (for the given sample properties and heating parameters) and $\Delta \bar{T}$ can be several Kelvin depending on the time after the heating pulse~\cite{Walter2011}.

In any case the Seebeck coefficient is not longer well defined and the current $I$ has to be considered, which is given by
\begin{equation}
\begin{aligned}
  I&(\bar{T},\Delta T,V_{\textrm{B}})= \frac{2e}{h} \int dE \: T(E,V_{\textrm{B}},\bar{T}) \ \cdot\\
	&\left( f(E,\mu_{\textrm{L}},\bar{T}-\frac{\Delta T}{2}) - f(E,\mu_{\textrm{R}},\bar{T}+\frac{\Delta T}{2})   \right ) \ \ \ ,
  \label{eq:TMS:current}
\end{aligned}	
\end{equation}

where $V_{\textrm{B}}=\frac{\mu_{\textrm{L}}-\mu_{\textrm{R}}}{e}$ is the applied bias voltage.
\begin{figure}[b!]
	\includegraphics[width=1 \linewidth]{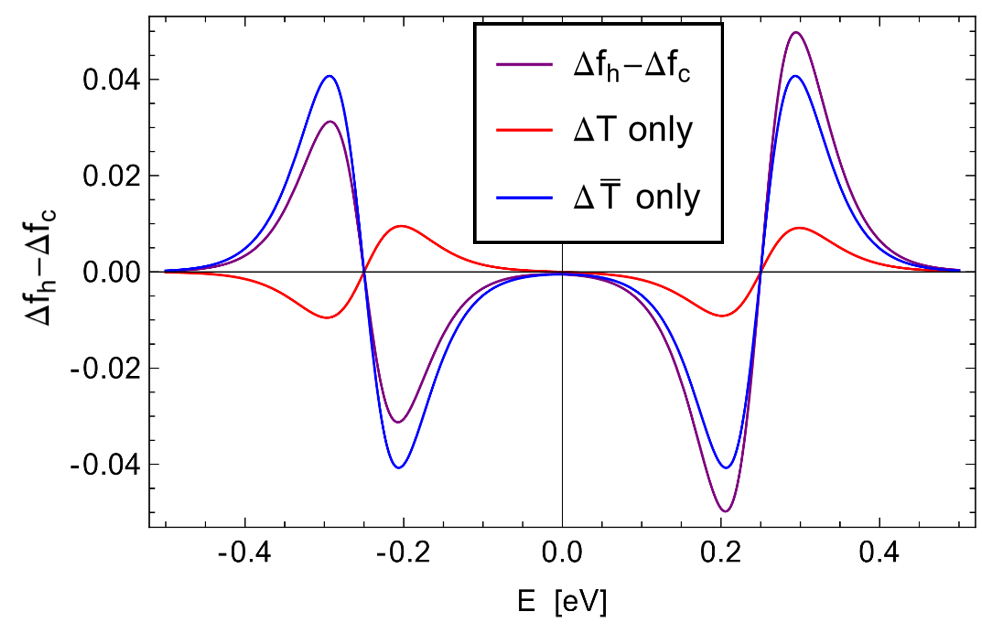}
	\caption{Differences of the occupation functions for the cold $f_{\textrm{c}}$ and hot $f_{\textrm{h}}$ case. In addition to the realistic case sketched in Fig.~\ref{fig:bTMS:temp} also the two limiting cases with only a temperature gradient ($\Delta \bar{T}=0$) and only a change of the mean temperature ($\Delta T=0$) are shown.}
	\label{fig:TMS:delta_f}
\end{figure}
Here, a cold current $I_{\textrm{c}}$ can be defined without any heating by using the occupation function
\begin{equation}
  \Delta f_{\textrm{c}}= f(E,\mu_{\textrm{L}},\bar{T})-f(E,\mu_{\textrm{R}},\bar{T}) \ \ \ .
\label{eq:fc}
\end{equation}

In the same way a hot current $I_{\textrm{h}}$ with heating is introduced by using the occupation function
\begin{equation}
\Delta f_{\textrm{h}}= f(E,\mu_{\textrm{L}},T_{\textrm{R}})- f(E,\mu_{\textrm{R}},T_{\textrm{L}}) \ \ \ .
\label{eq:fh}
\end{equation}

In the experiment one measures the difference between both currents
\begin{equation}
  \Delta I=I_{\textrm{h}}-I_{\textrm{c}}=\frac{2e}{h} \int dE \: T(E,V_{\textrm{B}},\bar{T}) \left ( \Delta f_{\textrm{h}} - \Delta f_{\textrm{c}}   \right ) \ \ \ .
  \label{eq:TMS:diffcurrent}
\end{equation}

In order to understand, which states are contributing to the detected current, $\Delta f_{\textrm{c}}- \Delta f_{\textrm{h}}$ should be analyzed. The principle is sketched in Fig.~\ref{fig:TMS:delta_f}. In all cases one gets indeed contributions from other states but they are from the lower and the upper limit of the bias voltage window. In the case of a pure temperature gradient ($\Delta \bar{T}=0$) the difference in the occupation function is anti-symmetric. The latter is symmetric in the case of a pure increase of the average temperature ($\Delta T=0$). In reality it is a mixture of both cases. For example, in the experiment by Boehnke et al.~\cite{Boehnke2015} finite element modeling shows that $\Delta \bar{T} >> \Delta T$ as already mentioned above.
\begin{figure}[b!]
	\includegraphics[width=1 \linewidth]{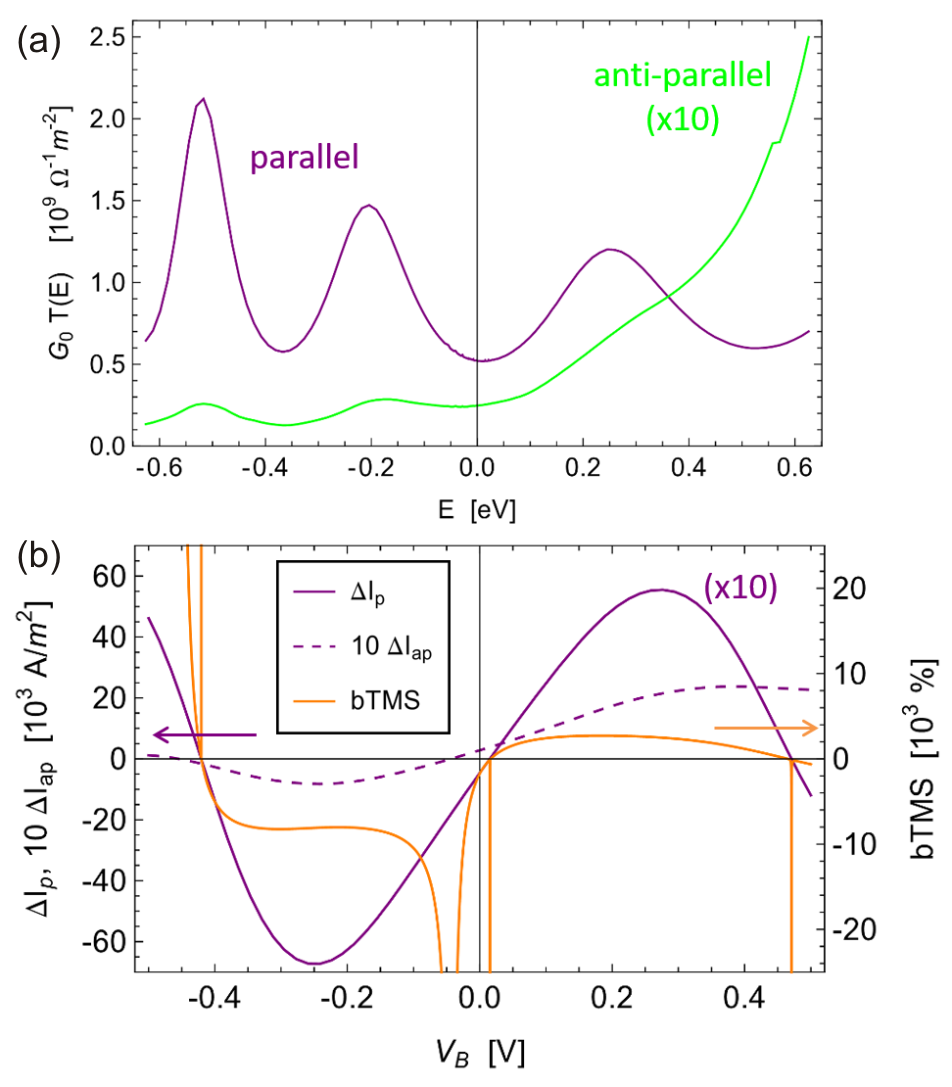}
	\caption{(a) Transmission functions for P and AP case of a Fe$_{0.7}$Co$_{0.3}$/8MgO/Fe$_{0.7}$Co$_{0.3}$ MTJ. (b) Corresponding currents (Eq.~(\ref{eq:TMS:diffcurrent})) and bTMS (Eq.~(\ref{eq:TMS:bTMS})) for $\bar{T}=300\,\textrm{K}$, $\Delta T=1\,\textrm{K}$, and $\Delta \bar{T}=1\,\textrm{K}$. }
	\label{fig:bTMS:cases}
\end{figure}
\begin{figure*}[t!]
\centering
\includegraphics[width=1 \linewidth]{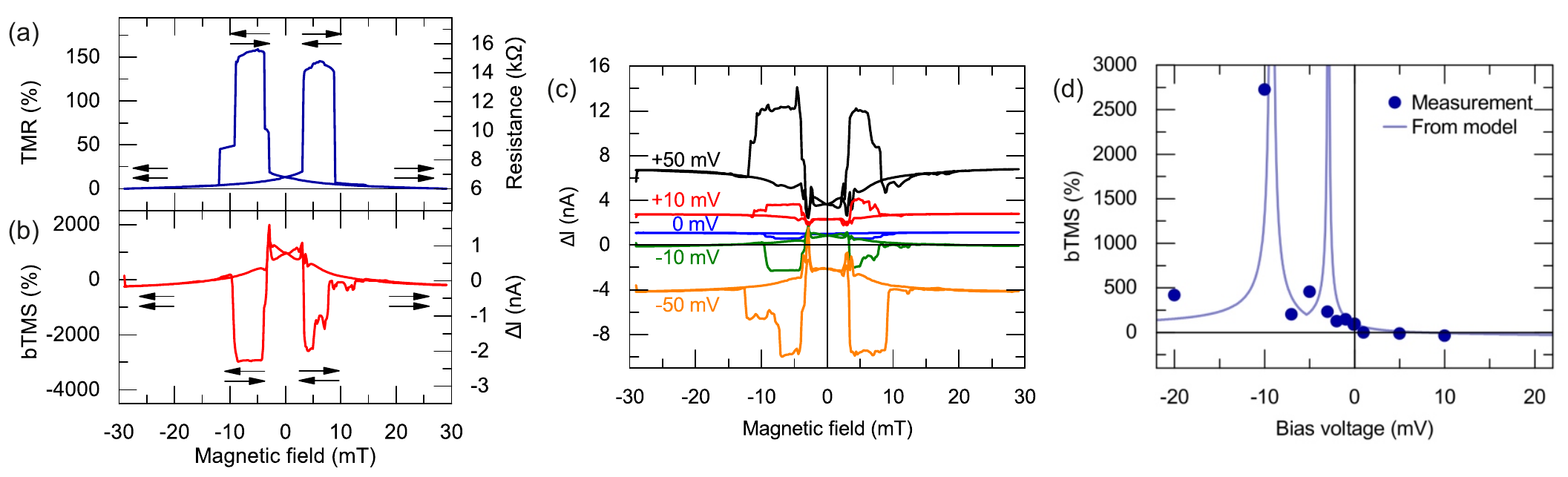}
\caption{Experimentally determined bias-enhanced TMS. (a) TMR ratio and resistance against magnetic field. (b) bTMS ratio and measured current signal for a bias voltage of $210~\textrm{mV}$ and a laser power of $150~\textrm{mW}$. The resulting effect ratio reaches almost -3000\%. (c) Variation of bias voltage for the bTMS. (d) Comparison to theoretical calculations. The experimental values (dots) are derived from (c). The solid line represents the calculation that clearly shows several divergences attributed to the vanishing $\Delta I$ in only one magnetic state of the MTJ resulting in such high effect ratios. Figures taken from Ref.~\cite{Boehnke2015}.}
\label{fig:exp-bTMS}
\end{figure*}

In order to get an idea about this effect Fig.~\ref{fig:bTMS:cases}(a) shows a realistic transmission function. From that one can calculate different scenarios using Eq.~(\ref{eq:TMS:diffcurrent}) for P and AP alignment. The bias-enhanced TMS can be defined by
\begin{equation}
  \textrm{bTMS}=\frac{\Delta I_{\textrm{P}}- \Delta I_{\textrm{AP}}}{\min (\left | \Delta I_{\textrm{P}} \right |, \left | \Delta I_{\textrm{AP}} \right |)} \ .
\label{eq:TMS:bTMS}
\end{equation}
The bTMS is shown for one scenario in Fig.~\ref{fig:bTMS:cases}(b). It is obvious that the bTMS can be easily tuned by the applied bias voltage. All this discussion neglects the influence of bias voltage and temperature on the transmission function. Therefore, this is a rather rough estimate and just gives an idea about the effect. 

Corresponding measurements for the case that $\Delta \bar{T} >> \Delta T$ are given by Boehnke et al.~\cite{Boehnke2015} using laser heating. As presented in Fig.~\ref{fig:exp-bTMS}(b), the bTMS can reach values up to 3000\%, although the TMR is not that high (Fig.~\ref{fig:exp-bTMS}(a)). By varying the bias voltage the $\Delta I$ can reach zero for one magnetic configuration (see Fig.~\ref{fig:exp-bTMS}(c)). This zero crossing of one Seebeck coefficient in the denominator of the definition of Eq.~(\ref{eq:TMS:bTMS}) explains the high TMS ratios as, e.g., presented in Fig.~\ref{fig:exp-bTMS}(b). These divergences are also observable in the theoretic calculations and can be compared with the experimental values (Fig.~\ref{fig:exp-bTMS}(d)). For practical applications the absolute voltage value of $S_{\textrm{P}}$ or $S_{\textrm{AP}}$ has to be kept at a reasonable signal amplitude at the same time in order to exploit these large effects with a low noise level.  

\subsection{Variation of electrode material}
\label{sec:variation-material}

\begin{figure}[b!]
	\includegraphics[width=1 \linewidth]{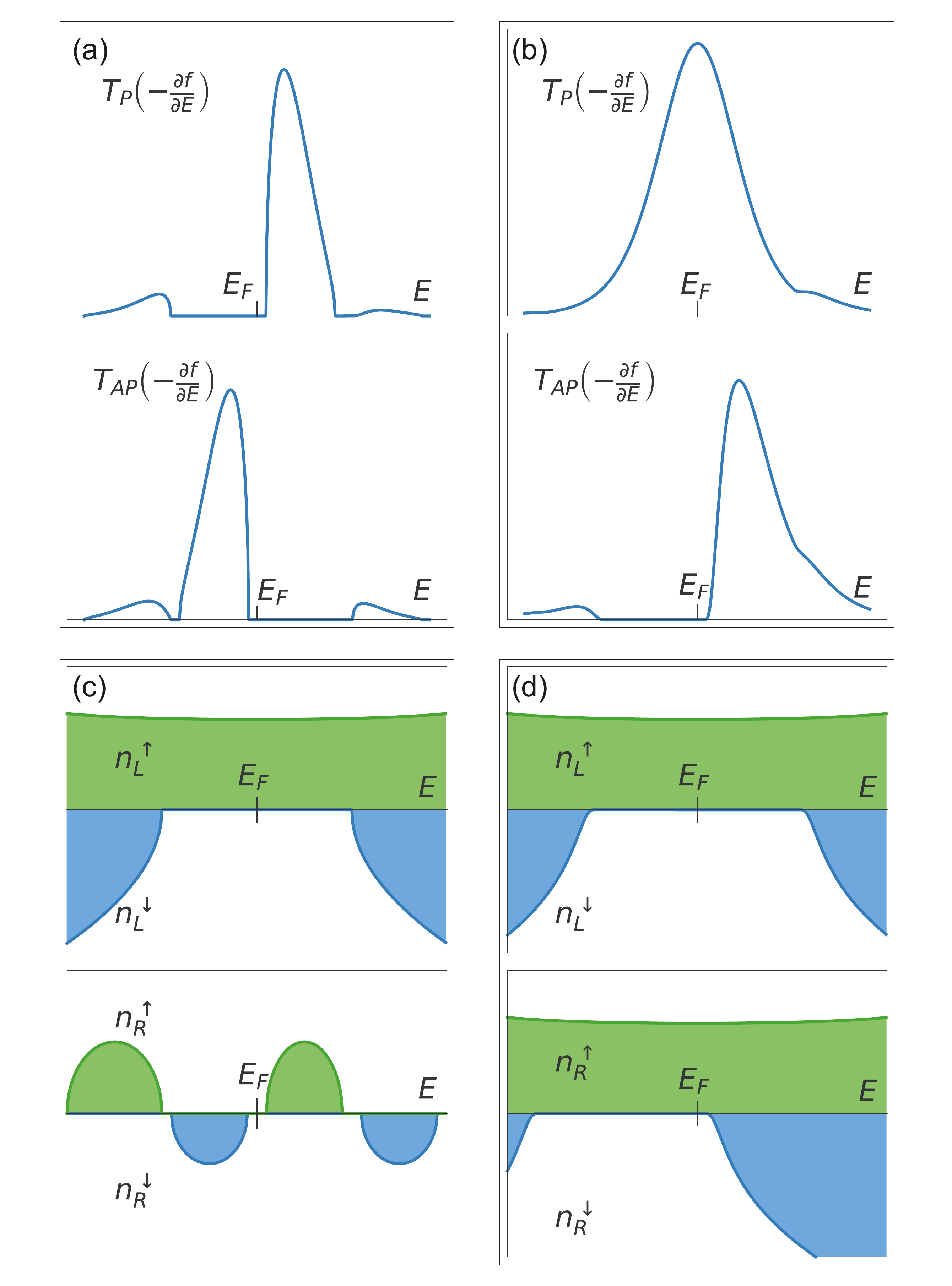}
	\caption{Ideal integrands~(see Eq.~(\ref{eq:integrand})) at some temperature for (a) different sign of $S_{\textrm{P}}$ and $S_{\textrm{AP}}$ and (b) for a very high TMS ratio with a vanishing $S_{\textrm{P}}$ and a large $S_{\textrm{AP}}$. Following the Julli$\acute{\textrm{e}}$re model~(see Eq.~(\ref{eq:Julliere})) the panels (c) and (d) show corresponding DOS that will lead to the desired transmission functions.}
	\label{fig:TMS:best_mat}
\end{figure}

The TMS ratios observed so far do not extend a few 10\%, if we do not consider any intrinsic TMS and the bTMS. In order to increase the TMS by the choice of proper materials, one should take into account the dependence of the TMS on the asymmetry of the spin-split DOS with respect to the Fermi energy. Thereby, one has to consider two different cases. One goal might be to have large Seebeck values for parallel and anti-parallel alignment but with opposite sign. Such transmission functions are shown in Fig.~\ref{fig:TMS:best_mat}(a). In this case the TMS ratio as defined by Eq.~(\ref{eq:def:TMS}) might be not that high.

Therefore, another idea is to have a very small Seebeck coefficient for one alignment and a large one for the other alignment. This situation is shown in Fig.~\ref{fig:TMS:best_mat}(b). One can discuss how an MTJ with such transmission functions should be designed. A simple way to answer this question is to apply the Julli$\acute{\textrm{e}}$re model~\cite{Julliere1975}, which connects in a simple way the available density of states $n$ to the transmission function via
\begin{equation}
\begin{aligned}
  T_{\textrm{P}} \propto & n_{\textrm{L}^\uparrow} n_{\textrm{R}^\uparrow} + n_{\textrm{L}^\downarrow} n_{\textrm{R}^\downarrow} \ \ \ ,\\
  T_{\textrm{AP}} \propto & n_{\textrm{L}^\uparrow} n_{\textrm{R}^\downarrow} + n_{\textrm{L}^\downarrow} n_{\textrm{R}^\uparrow} \ \ \ ,
	\end{aligned}
  \label{eq:Julliere}
\end{equation}
where $L$ and $R$ again refers to the material above and below the barrier, respectively. The Julli$\acute{\textrm{e}}$re model is valid only for incoherent tunneling but it gives some clue of what kind of material is necessary. Applying the Julli$\acute{\textrm{e}}$re model to the transmission function shown in Fig.~\ref{fig:TMS:best_mat}(a) leads to the finding that one magnetic layer has to be a half-metal whereas the other one could be a magnetic semiconductor with properly aligned bands. This is shown in Fig.~\ref{fig:TMS:best_mat}(c).

For the other case a possible solution is to have one half-metallic magnetic layer and a second half-metallic layer with properly aligned bands that show a peak in one spin channel close to the Fermi level. Indeed such a density of states is common for certain Heusler compounds~\cite{Meinert2012,Geisler2015,Boehnke2017}. A similar Heusler on the other side of the barrier would not get the desired behavior, because the two peaks would lead to a asymmetric transmission function in the AP case and, thus, to a non-vanishing Seebeck coefficient. In the case of the epitaxial MgO based MTJs it is very simple to get the half-metallic properties, since the MgO/Fe or MgO/Fe-Co interface acts for the transport properties like a half-metal~\cite{Butler2001,Heiliger2006a}. The reason is that MgO filters states with $\Delta_1$ symmetry and such states are only present in one spin channel of Fe.

\begin{figure}[t!]
\centering
\includegraphics[width=1 \linewidth]{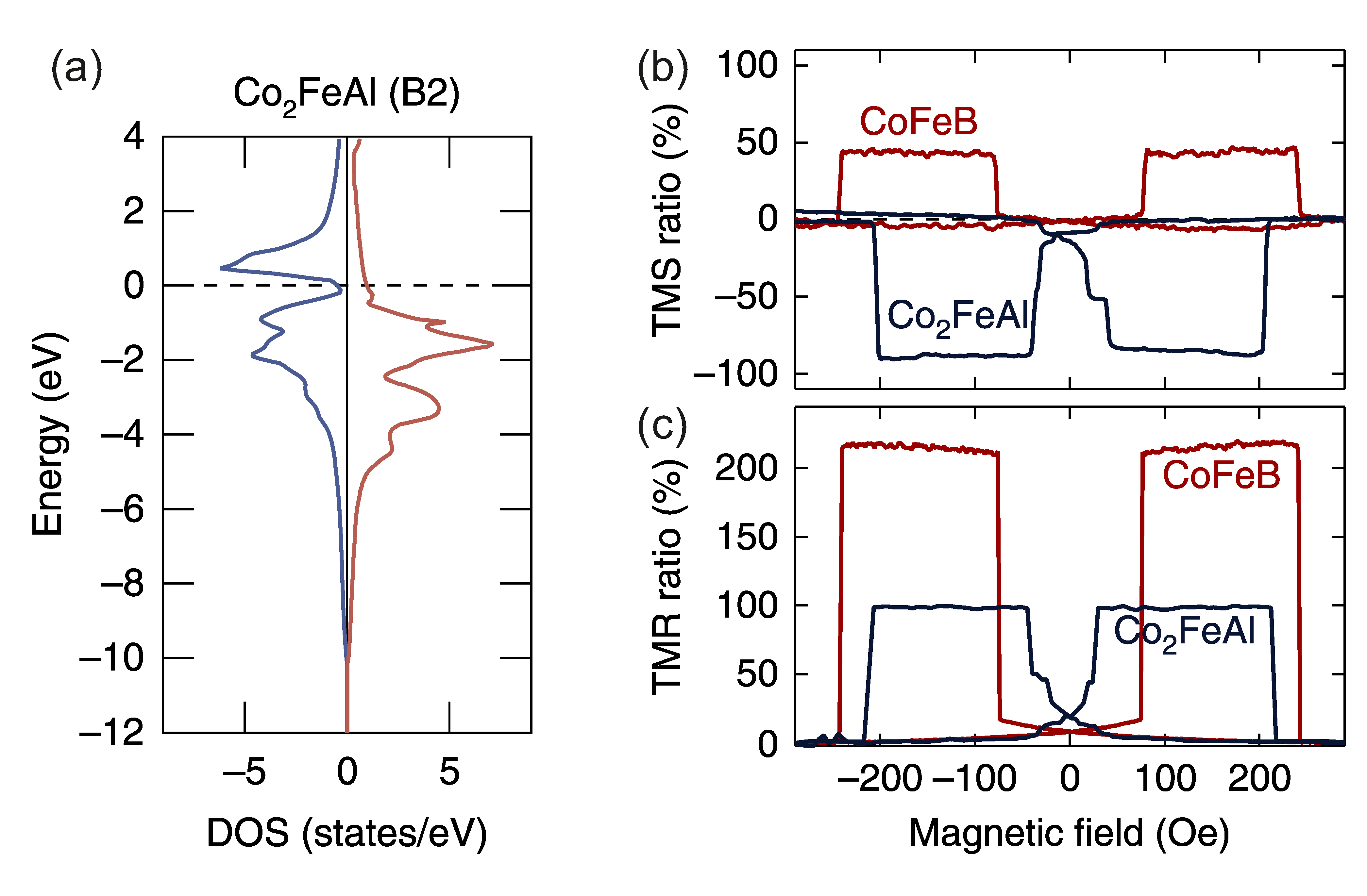}
\caption{TMS for MTJs with the half-metallic Heusler compound Co$_2$FeAl as electrode material. (a) DOS of Co$_2$FeAl in the B2 structure obtained by DFT calculations. The Fermi level is located at the edge of the band gap resulting in a large asymmetry of the DOS around the Fermi level and, thus, supporting a large TMS. (b) Comparison of the laser-heating induced TMS ratio between MTJs with Co-Fe-B and with Co$_2$FeAl electrode. (c) Comparison of the TMR ratio between MTJs with Co-Fe-B and with Co$_2$FeAl electrode. While the Co$_2$FeAl MTJs have much larger TMS ratios compared to the Co-Fe-B MTJs, the TMR ratio is reduced. Figures taken from Ref.~\cite{Boehnke2017}.}
\label{fig:TMS-Heusler}
\end{figure}

Therefore, in order to get a very high TMS ratio one should consider an MTJ of the form half-metallic Heusler compound/MgO/Fe-Co. Following this idea, Boehnke et al.~\cite{Boehnke2017} experimentally investigated MTJs with the half-metallic Heusler compounds Co$_2$FeAl and Co$_2$FeSi. As an example, we have picked the case of MTJs with one Co$_2$FeAl electrode to illustrate this successful way from material design to high effect values. The DOS of Co$_2$FeAl in the B2 structure calculated by DFT is shown in Fig.~\ref{fig:TMS-Heusler}(a).

One can clearly see that the Fermi level is located at the edge of the half-metallic band gap of Co$_2$FeAl in the B2 structure. This is also the case for Co$_2$FeSi in the L2$_1$ structure~\cite{Boehnke2017}. For L2$_1$ structured Co$_2$FeAl, the Fermi level is in the center of the band gap. That is why B2 structured Co$_2$FeAl is more likely to give a large TMS. Boehnke et al. prepared Co$_2$FeAl MTJs with Co-Fe-B counter electrode and compared TMS as well as TMR with MTJs having both electrodes made of Co-Fe-B. The results for laser-heating induced TMS are presented in Fig.~\ref{fig:TMS-Heusler}(b). While the TMS ratio of the Co-Fe-B MTJ stays below 50\%, the TMS ratio of the Co$_2$FeAl MTJ nearly exceeds -100\%. This result confirms the predictions made based on the calculated DOS and the position of the Fermi energy (Fig.~\ref{fig:TMS-Heusler}(a)) as well as the considerations made with respect to the desired transmission functions (Fig.~\ref{fig:TMS:best_mat}).

If we further compare the TMR ratios of the two types of MTJs (Fig.~\ref{fig:TMS-Heusler}(c)), we identify a smaller TMR for the Co$_2$FeAl MTJ compared to the Co-Fe-B MTJ. This again confirms the fact that TMS and TMR are not directly related. Boehnke et al. investigated multiple MTJs of each kind and also varied the Co-Fe composition in agreement with the theoretical predictions of Heiliger et al.~\cite{Heiliger2013a}. However, the same trend was observed for all samples: While large Seebeck coefficients in combination with large TMS values up to -120\% have been observed for the Heusler MTJs, the values are much smaller for the Co-Fe-B MTJs. This result is independent from the magnitude of the TMR ratio as systematically studied by Boehnke et al.~\cite{Boehnke2017}.

Further material variations have been calculated by Admin et al.~\cite{Admin2014} by taking advantage of materials with large spin-orbit coupling, such as Pt. They calculated a large tunnel anisotropy magneto-Seebeck effect in CoPt/MgO/Pt tunnel junctions of 175\%.

\subsection{Variation of barrier material and thickness}
\label{sec:variation-barrier}

The thickness of the insulating barrier and the choice of the barrier material play another important role. This has been systematically varied experimentally by Huebner et al.~\cite{Huebner2017} as well as B\"ohnert et al.~\cite{Boehnert2017a,Boehnert2018}. Beside MgO \cite{Walter2011,Liebing2011} and AlO$_{\textrm{x}}$~\cite{Lin2012} as barrier materials, the spinel ferrite MAO is an interesting candidate with fascinating properties. MAO has a lower lattice mismatch with standard ferromagnetic electrodes (Fe, CoFe, CoFeB, etc.) of 1\% compared to MgO that has between 3\% and 5\%~\cite{Miura2012}. This low value can be even more reduced by growing MgAl$_2$O$_{\textrm{x}}$ with reduced oxygen content via molecular beam epitaxy~\cite{Tao2015}. As a tunnel barrier, MAO exhibits a similar symmetry filter effect as MgO~\cite{Zhang2012b}. However, experimental TMR ratios obtained so far are still below the values for MgO-based MTJs~\cite{Sukegawa2010,Tao2014,Scheike2016}, while magnetization switching by spin-transfer torque has been demonstrated~\cite{Sukegawa2013} and MgAl$_2$O$_{\textrm{x}}$ double-barrier systems with pronounced resonant tunneling features in quantum well structures have been realized~\cite{Tao2015}.
\begin{figure}[b!]
\centering
\includegraphics[width=1 \linewidth]{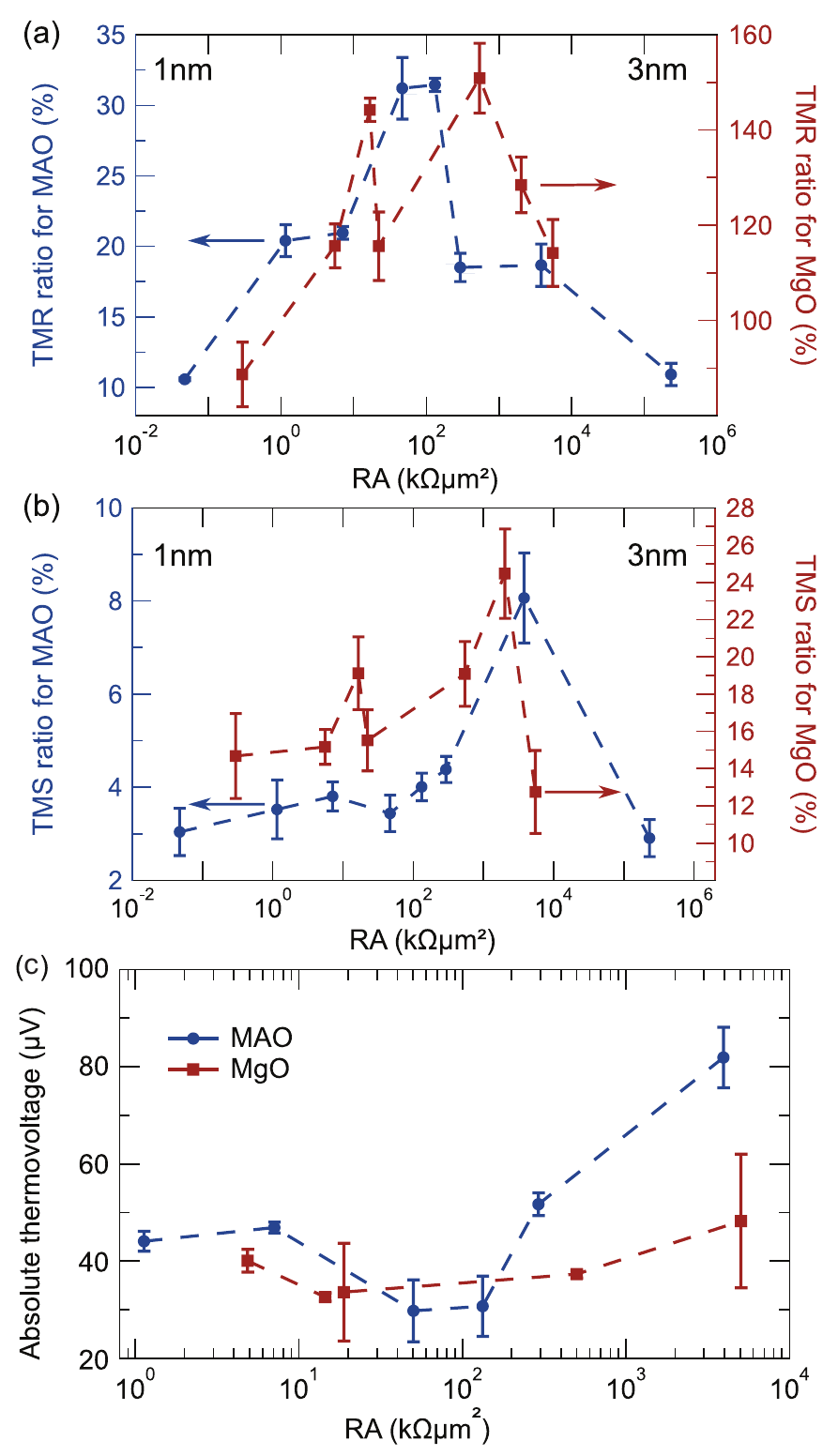}
\caption{Variation of the barrier thickness from $1\,\textrm{nm}$ to $3\,\textrm{nm}$ which affects the resistance area product $R\,A$ for MAO (left axis, blue circles) and MgO (right axis, red squares) barrier MTJs. (a) Comparison of the averaged TMR ratios over all measured elements plotted against $R\,A$. (b) Comparison of the laser-heating induced TMS ratios (averaged over all measured elements) plotted against $R\,A$. (c) Measured absolute thermopower with a laser power of $150\,\textrm{mW}$ of the MTJs with an area of $6\pi\mu\textrm{m}^2$ plotted against $R\,A$. Figures taken from Ref.~\cite{Huebner2017}.}
\label{fig:TMS-thickness-MAO}
\end{figure}

In Fig.~\ref{fig:TMS-thickness-MAO}, the TMR and TMS ratios together with the thermovoltage for MgO and MAO barrier MTJs with Co-Fe-B electrodes are plotted against the resistance area product $R\,A$ that scales with the barrier thickness \cite{Huebner2017} are summarized. The barrier thicknesses from $1\,\textrm{nm}$ to $3\,\textrm{nm}$ result in resistance area products in the range of $10^{-1}$ to $10^{5}\,\textrm{k}\Omega\mu\textrm{m}^2$. For each barrier thickness several MTJs are characterized and averaged. The TMR is maximal at around a nominal barrier thickness of 2 nm. Values up to 30\% for MAO based MTJs with $R\,A=100\,\textrm{k}\Omega\mu\textrm{m}^2$ and 150\% for MgO based MTJs with $R\,A=1000\,\textrm{k}\Omega\mu\textrm{m}^2$ can be observed (Fig.~\ref{fig:TMS-thickness-MAO}(a)).

If the TMS ratios between MAO and MgO based MTJs are compared, the TMS is again larger for MgO based MTJs (Fig.~\ref{fig:TMS-thickness-MAO}(b). The maximum is observed for both types of MTJs at a nominal barrier thickness of $2.6\,\textrm{nm}$ ($R\,A$ between $1000\,\textrm{k}\Omega\mu\textrm{m}^2$ and $10^4\,\textrm{k}\Omega\mu\textrm{m}^2$). Experiments and theory do not agree here, since theory predicts an increasing TMS ratio when going down from ten monolayers (MLs) (2\%) to six monolayers (10\%) of MgO ($1\,\textrm{ML}=2.1\,\textrm{\AA})$~\cite{Heiliger2013a}. However, the interface structure can be quite different in the experiment compared to the theory, where it is assumed to be ordered perfectly. Further decrease of TMR and TMS values for thinner barriers is due to pinholes that become more probable for smaller thicknesses and reduce the effect amplitudes. Since the TMS is sensitive to the interface structure~\cite{Czerner2012} and barrier properties, this can already lead to discrepancies between theory and experiment. 

Beside large TMS ratios, a huge thermopower is desirable. Therefore, the thermopower is plotted against the barrier thickness, i.e. $R\,A$ (Fig.~\ref{fig:TMS-thickness-MAO}(c)), for a laser power of $150\,\textrm{mW}$ and MTJs with an area of $6\pi\mu\textrm{m}^2$. Although the TMS is smaller in MAO based MTJs compared to MgO based ones, the thermopower is larger for the MAO barrier throughout most of the thicknesses tested. The highest value has been found for the thickest MAO barrier of $2.6\,\textrm{nm}$. The two samples with a nominal MAO thickness of $1.8\,\textrm{nm}$ and $2.0\,\textrm{nm}$ have been fabricated and measured separately under different experimental conditions which could explain the reduced thermopower values as discussed in Ref.~\cite{Huebner2017}. The thermopower has been taken after subtracting the lead contribution determined after electric breakdown of the MTJ by applying $3\,\textrm{V}$ to the junction.

In addition, B\"ohnert et al.~\cite{Boehnert2017a,Boehnert2018} investigated electric-heater induced TMS in MTJs with Co-Fe-B electrodes and an MgO barrier wedge with varied thicknesses from $1.2\,\textrm{nm}$ to $1.6\,\textrm{nm}$ related to $0.7\,\textrm{k}\Omega\mu\textrm{m}^2$ and $55\,\textrm{k}\Omega\mu\textrm{m}^2$, respectively. The obtained TMS values between 5\% and 35\% are in comparable order of magnitude as the results of Huebner et al.~\cite{Huebner2017}. Theoretical studies on MTJs with varying thickness of an aluminium oxide barrier has been conducted by L\'{o}pez-Mon\'{i}s et al.~\cite{Lopez-Monis2014}. However, experimental investigations for this kind of MTJs with various thicknesses of the aluminium oxide barrier are still missing.

\section{Further aspects and applications}
\label{aspects}

\subsection{Angular dependence of the TMS}
\label{sec:angular-dep}
\begin{figure}[b!]
\centering
	\includegraphics[width=1 \linewidth]{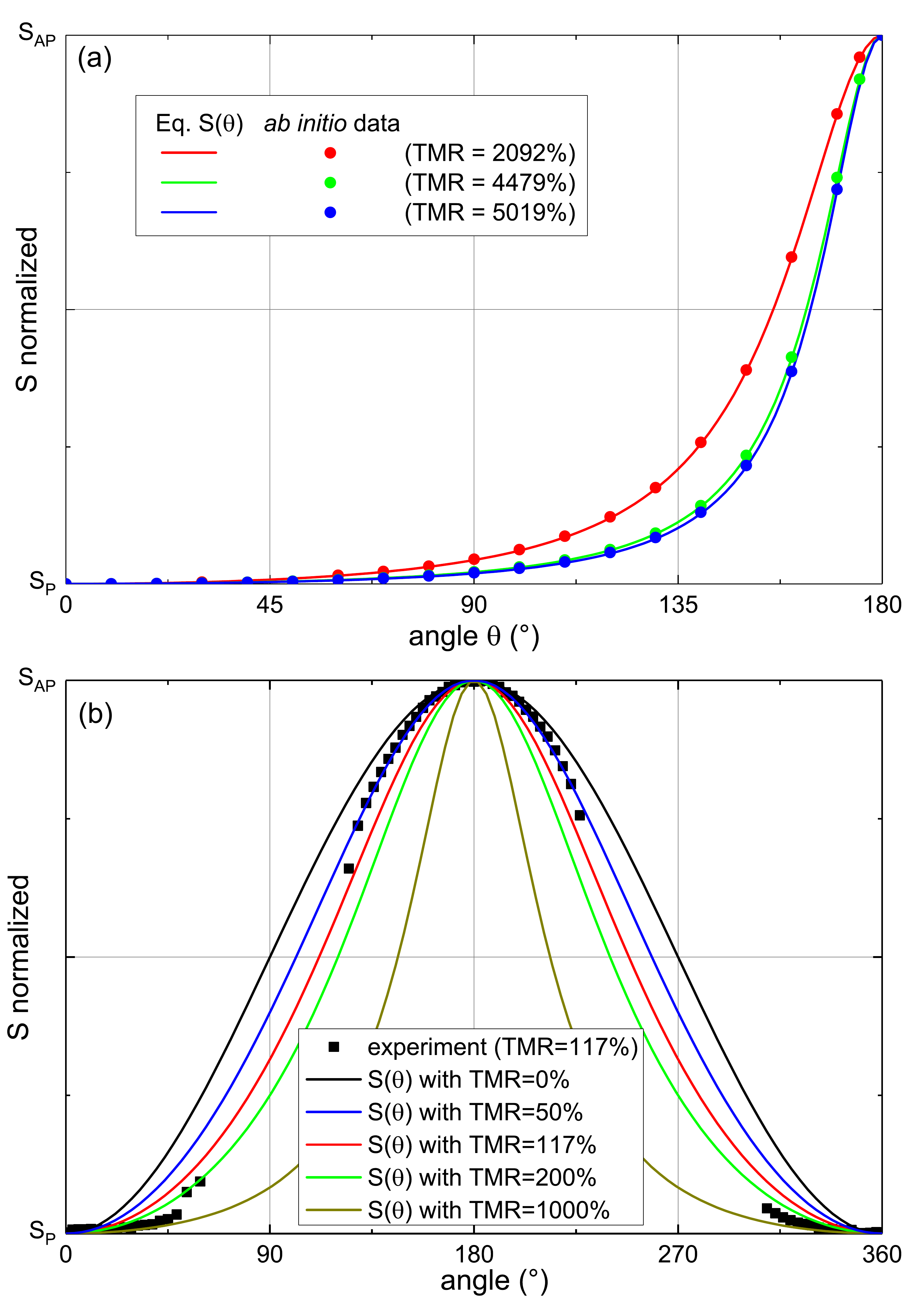}
	\caption{Angular dependence of the TMS. (a) Seebeck coefficients obtained by \textit{ab initio} calculations~\cite{Czerner2011,Walter2011,Heiliger2013a}. The symbols are the angles where the calculations have been done and the solid lines are Eq.~(\ref{eqn:Stheta}). (b) Experimental data of the angular dependence of the Seebeck coefficient compared to Eq.~(\ref{eqn:Stheta}) for different TMR ratios. Figures taken from Ref.~\cite{Heiliger2013b}.}
	\label{fig:TMS:angle}
\end{figure}
Up to now we discussed the P and AP alignment of the magnetic moments. However, the magnetization of the two ferromagnetic layers can be oriented in any direction to each other. For the case of the TMR or the STT this gives basically a simple cosine dependence~\cite{Slonczewski1989,Heiliger2008,Jaffres2001}. For the TMS this can be different~\cite{Heiliger2013b}. Figure~\ref{fig:TMS:angle}(a) shows the calculated normalized angular dependence for three different MTJs. The main difference between these junctions is the TMR ratio. It can be shown that the angular dependence of the Seebeck coefficient $S$ follows~\cite{Heiliger2013b}
\begin{equation}
\footnotesize{S(\theta)=\frac{S_{\textrm{\tiny{P}}}\cdot\textrm{TMR}+S_{\textrm{\tiny{P}}}+S_{\textrm{\tiny{AP}}}+(S_{\textrm{\tiny{P}}}\cdot\textrm{TMR}+S_{\textrm{\tiny{P}}}-S_{\textrm{\tiny{AP}}})\cdot\cos(\theta)}{\textrm{TMR}+2+\textrm{TMR}\cdot\cos(\theta)}\ .}\\
\label{eqn:Stheta}
\end{equation}

From that it becomes clear that in the limiting case of vanishing TMR the TMS obeys a cosine dependence. This explains also different results between theory and experiments, because in experiments TMR ratios are typically between 100\% and 300\% whereas in theory they are typically far above 1000\%. In Fig.~\ref{fig:TMS:angle}(b) we show experimental results compared to Eq.~(\ref{eqn:Stheta}) for different TMR ratios. One obtains fairly good agreement to Eq.~(\ref{eqn:Stheta}) by plugging in the TMR ratio of the experiment. The results show that although the TMS and TMR ratios are independent they are connected via the angular dependence. In particular, the TMR ratio can be fitted by applying Eq.~(\ref{eqn:Stheta}) to the experimental data.

\subsection{Temperature drop across a tunnel barrier}
\label{sec:thermal-resistivity}

All the discussed experiments need a temperature gradient across an MTJ. Whereas in theory the transport parameters are in linear response, i.e. with vanishing temperature gradient, in experiment one has a basically unknown temperature drop across the barrier. However, this drop is important to calculate for example the Seebeck coefficient, because a thermovoltage is measured only. Recently, there are some experimental attempts to measure the temperature drop locally~\cite{Boehnert2017b,Yang2017}. But typically, finite element simulations are carried out in order to estimate the temperature drop~\cite{Walter2011,Liebing2011,Huebner2016}.
\begin{figure}[t!]
	\includegraphics[width=1 \linewidth]{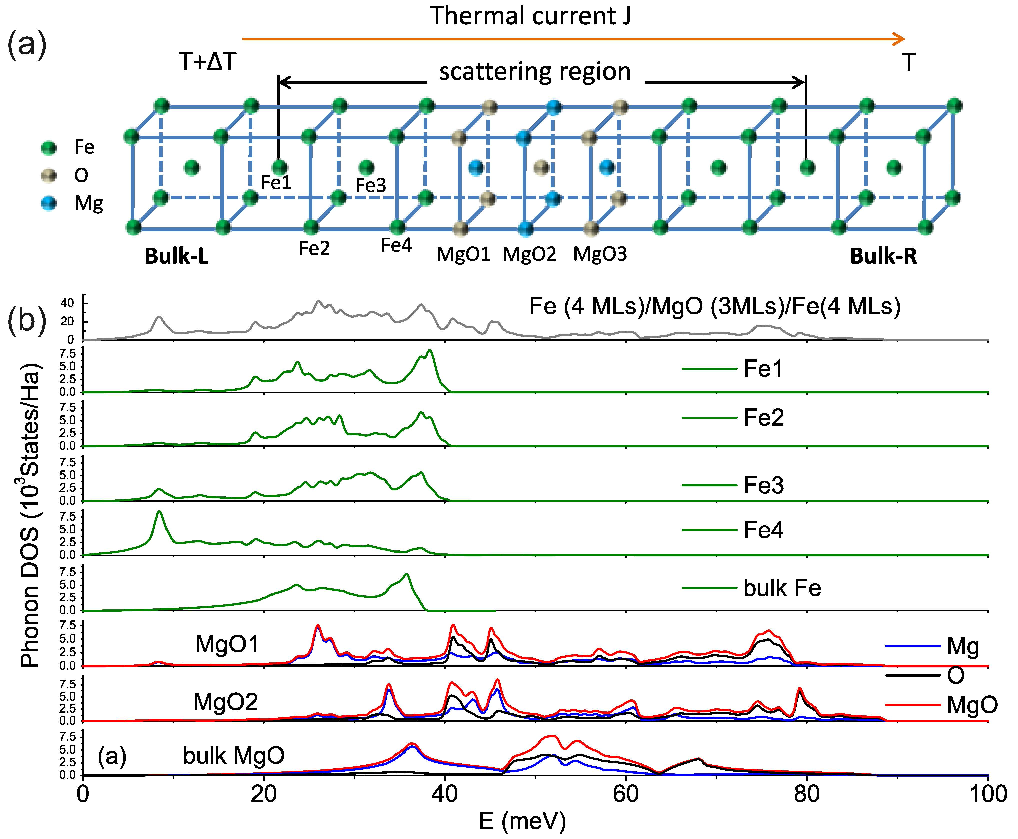}
	\caption{(a) The sketch of the Fe/MgO(3ML)/Fe MTJ for the calculation of phonon transmission function. The supercell used for the phonon calculations is marked as scattering region. (b) The corresponding total and projected phonon density of states (PDOS). In addition the bulk PDOS of Fe and MgO are shown. Figure taken from Ref.~\cite{Zhang2015}.}
	\label{fig:temp:pdos}
\end{figure}

One drawback of this simulation is that it assumes classical diffusion equations and transport properties of bulk, thin film, or interfaces as an input. However, in a nanostructure such as in an MTJ where the layers are only a few monolayers thick it is questionable if classical transport is valid and if even a temperature can be rigidly defined. In addition, any contribution of interface thermal resistances has been neglected so far in these simulations. Moreover, in a case of an MTJ the thermal transport through the metallic layers will be dominated by electrons whereas the transport through the insulator will be dominated by phonons.

Therefore, Zhang et al.~\cite{Zhang2015} calculated the phonon DOS (PDOS) for an MTJ with 3 monolayers MgO sandwiched between 4 monolayers of Fe using non-equilibrium Green's functions. The structure together with projected phonon density of states are shown in Fig.~\ref{fig:temp:pdos}. In addition, also the PDOS of bulk MgO and bulk Fe is shown. By comparing these two it is obvious that there is a huge mismatch. In particular, there is a cut-off energy of about $40\,\textrm{meV}$ given by Fe, and at the energy range from $20\,\textrm{meV}$ to $30\,\textrm{meV}$ where Fe has a rather high PDOS, MgO shows very low PDOS. Such a mismatch means that there is a very high thermal interface resistance.

\begin{figure}[b!]
	\includegraphics[width=1 \linewidth]{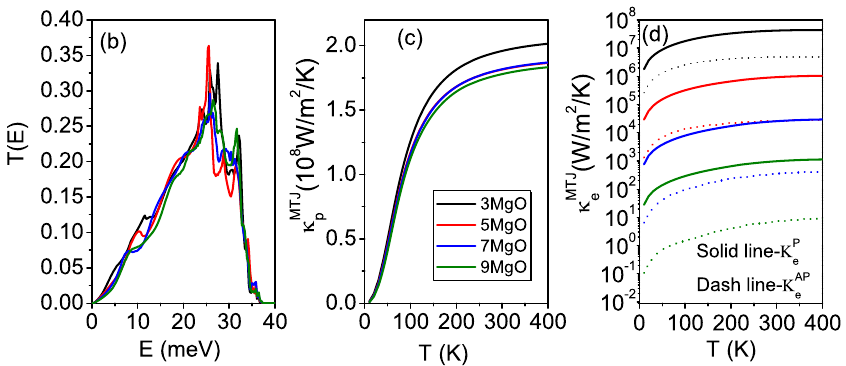}
	\caption{(a) Phonon transmission function, (b) phonon conductance $\kappa_p^{MTJ}$, and (c) electron thermal conductance of Fe/MgO($3-9\,\textrm{ML}$)/Fe as a function of temperature. Figure taken from Ref.~\cite{Zhang2015}.}
	\label{fig:temp:trans}
\end{figure}

This is reflected in the corresponding transmission function shown in Fig.~\ref{fig:temp:trans}(a). Thereby, the transmission function is almost independent of the MgO thickness. From the transmission function one can calculate the thermal conductance shown in Fig.~\ref{fig:temp:trans}(b). These values already show orders of magnitude smaller values than bulk MgO~\cite{Hofmeister2014} and even smaller than MgO thin films~\cite{Lee1995}. For comparison Fig.~\ref{fig:temp:trans}(c) shows the electron thermal conductance. As expected it decays exponentially with the increasing MgO thickness. Even at thin MgO barriers it is one order of magnitude smaller than the phonon thermal conductance.

\begin{figure}[t!]
	\includegraphics[width=1 \linewidth]{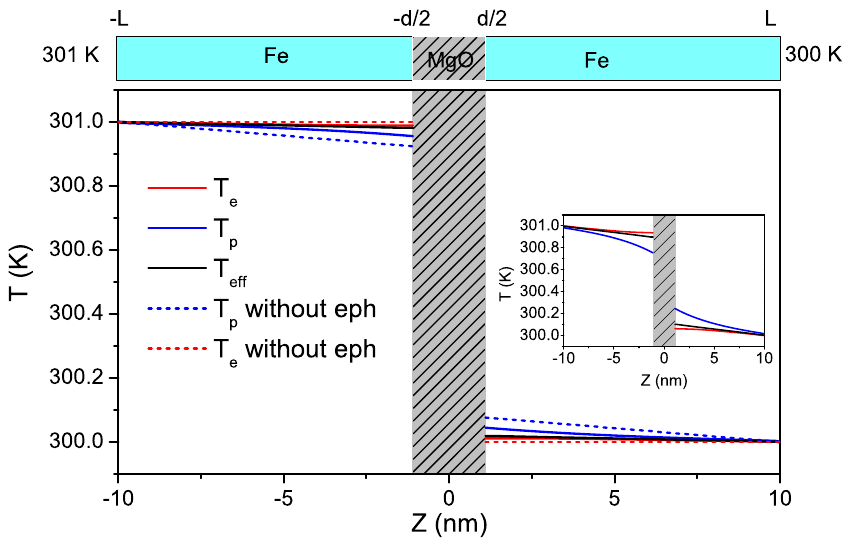}
	\caption{Temperature profile across a Fe/MgO(9 ML)/Fe MTJ. $T_{\textrm{e}}$, $T_{\textrm{p}}$, and $T_{\textrm{eff}}$ are the electron, phonon, and linear extrapolation temperatures, respectively. The inset shows the temperature profile for the fictitious case of a ten times larger $\kappa_{\textrm{p}}^{\textrm{MTJ}}$. Figure taken from Ref.~\cite{Zhang2015}.}
	\label{fig:temp:temp}
\end{figure}

In order to get the thermal conductance of the whole MTJ the energy balance equations can be solved for the electron and phonon temperature, $T_{\textrm{e}}$ and $T_{\textrm{p}}$,~\cite{Zhang2015,Majumdar2004}
\begin{equation}
\begin{aligned}
k_{\textrm{e}}^{\textrm{Fe}}\frac{d^2T_{\textrm{e}}(z)}{dz^2}-G_{\textrm{eph}}(T_{\textrm{e}}-T_{\textrm{p}}) &=0 \ \ \ , \\ 
k_{\textrm{p}}^{\textrm{Fe}}\frac{d^2T_{\textrm{p}}(z)}{dz^2}+G_{\textrm{eph}}(T_{\textrm{e}}-T_{\textrm{p}}) &=0 \ \ \ ,
\label{eq:temp:balance}
\end{aligned}
\end{equation}
where $z$ is the transport direction and $G_{\textrm{eph}}$ is the electron-phonon interaction factor~\cite{Lin2008}. $k_{\textrm{e}}^{\textrm{Fe}}$ and $k_{\textrm{p}}^{\textrm{Fe}}$ are the electron and phonon thermal conductivities in Fe, respectively. The results are shown in Fig.~\ref{fig:temp:temp}.

It becomes clear that there is a huge temperature drop at the MgO barrier, which was already expected from Fig.~\ref{fig:temp:trans}. Further, an imbalance of the electron and phonon temperature at the interface is visible. The reason is that the electrons get more reflected than the phonons at the interface and thus show a heat accumulation. The electron phonon coupling leads to an equilibration of $T_{\textrm{e}}$ and $T_{\textrm{p}}$ away from the interface. The inset of Fig.~\ref{fig:temp:temp} shows a fictitious case with an MTJ that have a 10 times larger thermal conductance. In this case the difference of electron and phonon temperature is further increased.
\begin{figure}[t!]
	\includegraphics[width=1 \linewidth]{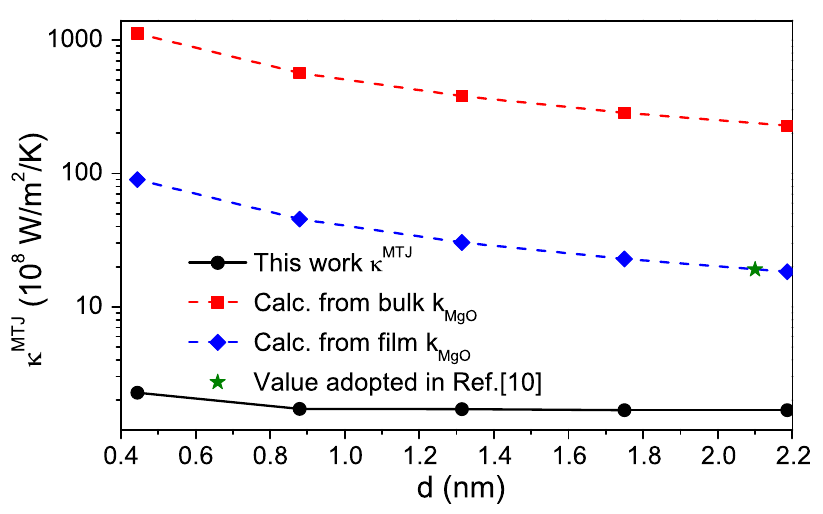}
	\caption{Thermal conductance of the MTJ as a function of barrier thickness.The results (black solid line) are compared to bulk (red dashed line) and thin film values (blue dashed line) of MgO. In addition, the value adopted in the finite element simulations of Ref.~\cite{Walter2011} is marked. Figure taken from Ref.~\cite{Zhang2015}.}
	\label{fig:temp:conductance}
\end{figure}

Although there is a non-equilibrium situation at the Fe/MgO interface one can estimate the total thermal conductance through the whole MgO stack~\cite{Zhang2015}. These values as a function of MgO thickness are given in Fig.~\ref{fig:temp:conductance} at room temperature and can be used as input parameters in finite element methods. 

In Fig.~\ref{fig:temp:conductance} these values are compared to values extracted from bulk MgO and thin films MgO. Going from bulk to thin films the thermal conductance is already decreased by one order of magnitude but it is again reduced by one order of magnitude going to the MTJ. Thus the values used in the past for finite element methods~\cite{Walter2011} lead to an underestimation of the temperature drop and thus to an overestimation of the Seebeck coefficient. This finding is confirmed by recent experiments~\cite{Huebner2018,Boehnert2017b}. 

\subsection{Three-dimensional thermal gradients and additional (magneto)thermoelectric effects}
\label{sec:ANE}
Additional (magneto)thermoelectric effects can occur in TMS experiments depending on the used MTJ materials and the heating technique. Here, we will discuss two examples. If Si is used as substrate for the MTJs, additional Seebeck voltages~\cite{Boehnke2013} and photocurrents~\cite{Xu2016} can be created in laser-heating induced TMS experiments depending on the spot size of the laser. Furthermore, unintended in-plane thermal gradients can end up in anomalous Nernst effect (ANE) contributions in the electrodes as systematically studied by Martens et al.~\cite{Martens2018}. Since the heat dissipation of the laser spot occurs in all spatial directions, additional (magneto)thermoelectric effects open new opportunities for the detection of the overall direction of the thermal gradient based on different heating scenarios and the control of the temperature gradient direction. 
\begin{figure}[b!]
	\includegraphics[width=1 \linewidth]{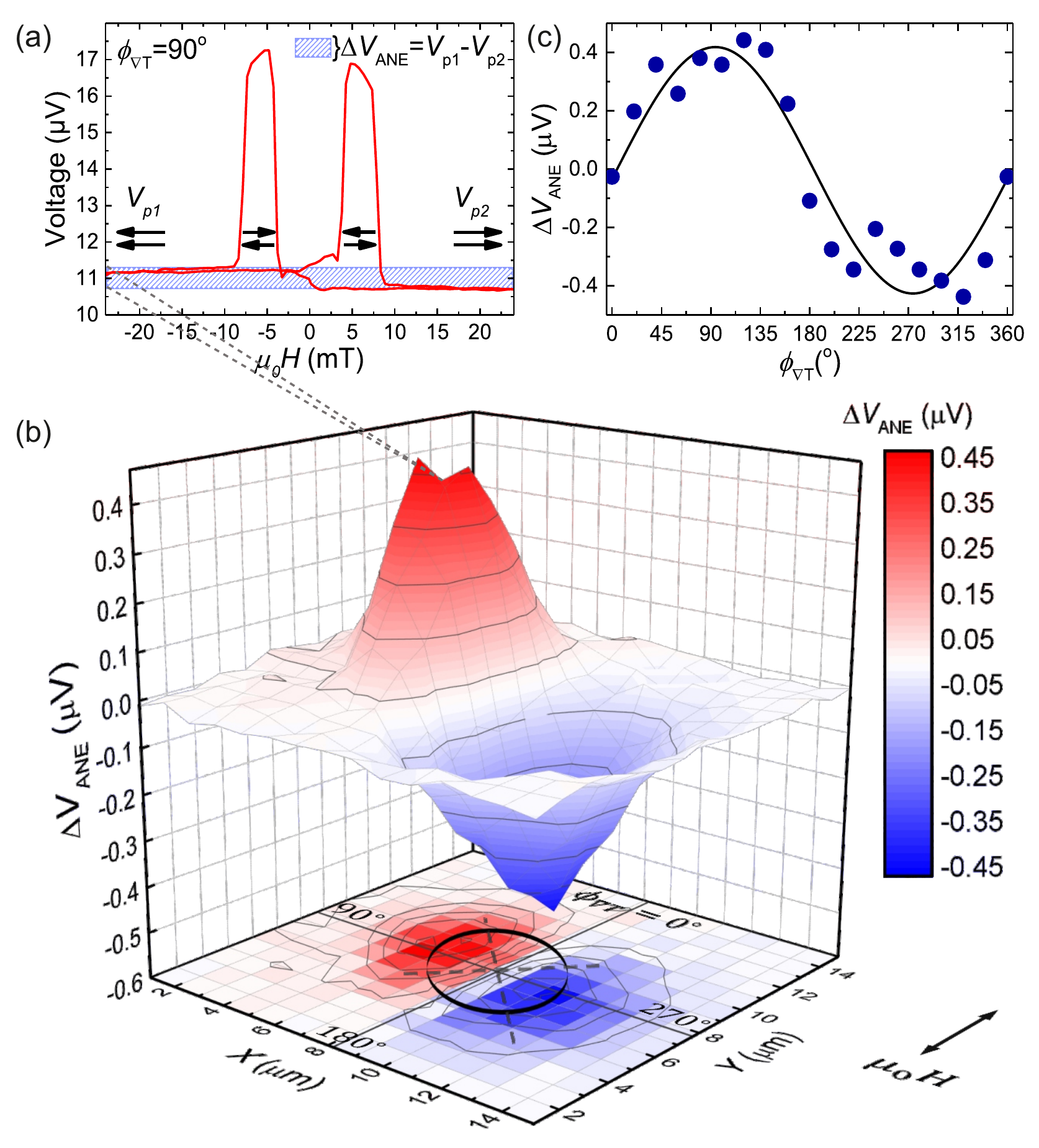}
	\caption{Identification of the ANE in TMS measurements. (a) TMS magnetic field loop with different offset values for opposite parallel magnetic alignments $V_{\textrm{p1}}$ and $V_{\textrm{p2}}$ indicated by the black arrows. The blue area marks the difference $\Delta V_{\textrm{ANE}}$ between the voltages $V_{\textrm{p1}}$ and $V_{\textrm{p2}}$. (b) The extracted $\Delta V_{\textrm{ANE}}$ values depending on the laser position. The whole plot is projected to the bottom of the graph with a black ellipse indicating the MTJ area. (c) The $\Delta V_{\textrm{ANE}}$ values along this ellipse are plotted against the angle $\phi_{\nabla\textrm{T}}$ of the temperature gradient describing the expected sinusoidal dependence of the ANE. Figures taken from Ref.~\cite{Martens2018}.}
	\label{fig:ANE-TMS}
\end{figure}

The lock-in detection of the laser-heating induced TMS experiments allows the time-resolved measurements of the TMS. Thus, any rising time of the thermovoltages and additional thermoelectric effects can be identified. By comparing the time-dependent TMS for MTJs on MgO and SiO$_2$/Si substrates, Boehnke et al. identified an additional Seebeck effect from the Si substrate that is not detectable in case of MgO~\cite{Boehnke2013}. While for MgO the thermovoltage increases within hundreds of $\mu\textrm{s}$ up to a plateau value, the thermovoltage of the Si samples overshoots this plateau value and decreases back down, probably due to additional capacitive couplings in the sample structure. Boehnke et al. developed a model circuit that considers the additional thermovoltage from the Si together with a capacitive coupling through the SiO$_2$ layer and could nicely describe the time-dependent experimental data. 

The effect of a Seebeck voltage from the substrate is only pronounced, if the laser spot is placed acentric on the MTJ~\cite{Boehnke2013}. This hints to the fact that additional in-plane thermal gradients induce the Seebeck effects of the substrate. As long as they are not canceled out (as it is the case for the acentric position of the laser spot), the Seebeck voltage of the substrate is contributing. Since MgO is insulating compared to the semiconducting Si, no additional Seebeck effect can be observed for MgO substrates independent from the position of the laser spot.

Beside this thermoelectric effect from in-plane thermal gradients, Martens et al. studied magnetothermoelectric effects, such as the ANE~\cite{Huang2011,Meier2013,Schmid2013}, for in-plane thermal gradients in the electrode material~\cite{Martens2018}. They are able to control the temperature gradient direction with respect to the magnetization alignment and extract the contribution of the ANE to the TMS signal. The measurements of TMS voltages in pseudo spin valves reveal a characteristic shift, (see Fig.~\ref{fig:ANE-TMS}(a)) for parallel magnetization alignment in opposite magnetic field directions.

The shift amplitude for each heating position is plotted in Fig.~\ref{fig:ANE-TMS}(b) in a three-dimensional plot and projected to the bottom of the graph with a black elliptically shaped contour line that indicates the MTJ position. The analysis of the extracted voltage shows a behavior that corresponds to the ANE. Especially the voltages $\Delta V_{\textrm{ANE}}$ extracted at heating positions at the MTJs edge, marked by the contour (see Fig.~\ref{fig:ANE-TMS}(c)) can be described by a sinusoidal dependence of the magnetization direction pointing to the typical angular dependence of the ANE~\cite{Martens2018}.

Future work could relate the individual effects to quantitative values that are reliable enough to compare the thermal gradient components of the TMS (out-of-plane component of $\nabla\textrm{T}$) and the ANE (in-plane component of $\nabla\textrm{T}$). Thus, one could receive the effective spatial direction of the thermal gradient and create a three-dimensional temperature gradient sensing device.\\

\subsection{Vacuum gap as tunnel barrier}
The TMS can be also observed on atomic scale if a vacuum gap is used as the tunnel barrier~\cite{Friesen2018a,Friesen2018b}. Stefan Krause's group from Hamburg University investigates STT and TMS on magnetic single and double layers using magnetic tips for scanning tunneling microscopy (STM). As depicted in Fig. 18(a), the tip is heated by a laser and, thus, creates the temperature gradient. The thermal expansion of the tip is used to determine quantitative numbers for the temperature difference. Due to this laser-heating induced thermal gradient the TMS can be detected as a thermovoltage and be mapped across the studied surface with atomic resolution for different magnetic moment alignments depending on the given sample.

Here, Friesen et al. have studied the TMS in Fe single and double layers grown on W(110) and Ir(111) substrates~\cite{Friesen2018b}. They clearly observe a linear dependence of the temperature difference between tip and surface on the magnetothermopower. Mapping of the local compensation bias voltage $U_c$ of the thermopower for zero DC tunnel current leads to the spatially-resolved Seebeck coefficient $S$. The compensation to zero tunnel current allows the exclusion of additional DC Joule heating effects. This measurement technique that is based on AC tunnel currents can be even applied for a vanishing temperature gradient, as discussed in Ref.~\cite{Friesen2018b}. In single layers of Fe on W(110) the magnetic moments are preferred in-plane aligned while for double layers the out-of-plane alignment is more favorable and the moments align in magnetic domains and domain walls, which is also observable in the mapping of the Seebeck coefficient.

\begin{figure}[t!]
	\includegraphics[width=1 \linewidth]{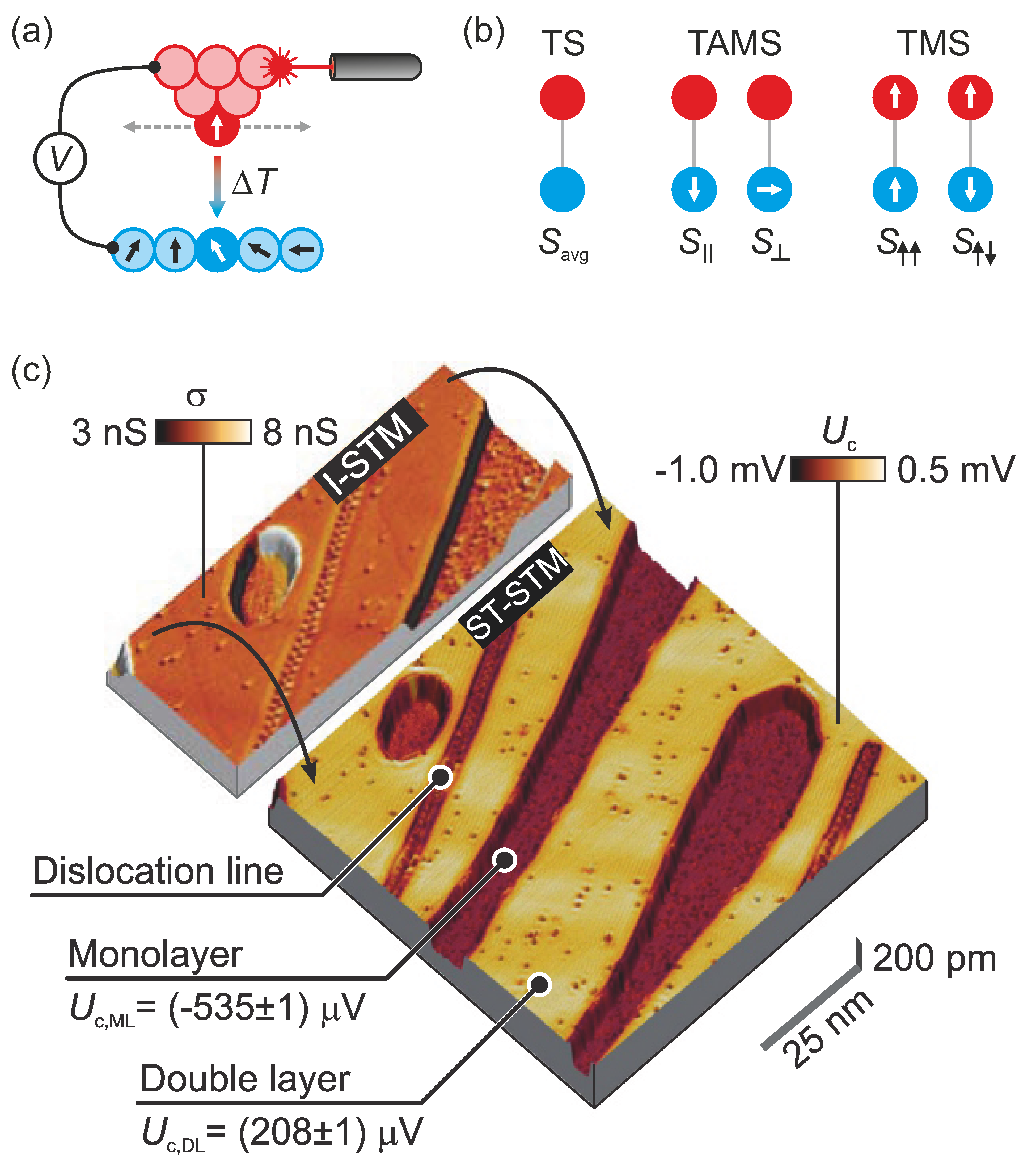}
	\caption{(a) Schematics of a single atom tunnel contact between a magnetic tip and a magnetic sample in an STM setup. The tip is heated by a laser beam, thereby creating a temperature difference $\Delta T$. The resulting thermovoltage $V$ across the junction is determined under compensated conditions. (b) Spatially-resolved determination of the Seebeck coefficient $S$ allows the component-resolved decomposition of spin-averaged tunneling Seebeck (TS), tunneling anisotropic magneto-Seebeck (TAMS) and TMS thermopower. Figures provided by Stefan Krause. (c) Comparison of standard tunnel current STM (I-STM) and the Seebeck tunneling mode of the STM with heated tip (ST-STM) obtained in compensated conditions. The images are taken on the same area on a Fe/W(110) sample at a temperature of $50\,\textrm{K}$. Single and double layer can be clearly identified. Figure taken from Ref.~\cite{Friesen2018a}.}
	\label{fig:STM-TMS}
\end{figure}

The Seebeck coefficients on the double layers can be described by a combination of spin-averaged Seebeck tunneling (ST), TMS and tunneling anisotropic magneto-Seebeck (TAMS) thermopower generated by spin-orbit coupling and by the different aligned magnetic moments within the magnetic domains and across the domain walls. The difference between three effects is shown in Fig.~\ref{fig:STM-TMS}(b) and they have been separated in the work of Friesen et al.~\cite{Friesen2018b}. As an example, the mapping of the spin-averaged ST response is shown in Fig.~\cite{Friesen2018a} with the $U_c$ values of the single and double layers for a Fe/W(110) sample. A standard tunnel current STM (I-STM) image of the same area is presented for comparison. The mapping of the magnetic response and additional results such as the mapping of $S$ for the Skyrmion material Fe/Ir(111) can be found in Ref.~\cite{Friesen2018a}. Generally, thermovoltages in STM experiments show fine contrast and features of electronic states also in nonmagnetic samples because of their different dependence on the local density of states as compared to an electrically biased tunneling.

\section{Conclusion and outlook}

The tunnel magneto-Seebeck effect (TMS) in magnetic tunnel junctions (MTJs) is a fascinating spin caloritronic effect that provides the possibility to obtain thermovoltages depending on the magnetic configuration of the MTJ and, thus, to use heat for future elements of spintronic circuits. We have reviewed the successful way of the TMS from first detection by laser- and electric heating in Co-Fe-B/MgO/Co-Fe-B MTJs~\cite{Walter2011,Liebing2011} over material optimization aspects to open questions that still have to be solved.

The experimental equipment and theoretical implementations are already on a very high level, so that ab initio calculations of band structure and transmission function nicely agree with the features that are observed in experiments. This scientific tango between theory and experiment~\cite{Sinova2014} led, for example, to the prediction and experimental confirmation of high TMS ratios in MTJs with half-metallic Heusler electrodes~\cite{Boehnke2017}. We further presented the increase of thermovoltages for the use of MgAl$_2$O$_4$(MAO) barriers with an optimized barrier thickness of $2.6\,\textrm{nm}$~\cite{Huebner2017}. Future studies should be made for Heusler/MAO/Co-Fe-B MTJs to adjust both high TMS ratios and large thermovoltages.

Further aspects that have to be addressed are the disentanglement of any intrinsic TMS effect induced by the tunnel current itself from the characteristic properties of the barrier shape that influence the V-I curve in a similar way~\cite{Huebner2016}. The role of the barrier thermal conductivity and additional thermal interface conductivities between barrier and electrodes are not yet fully explored. Due to the theoretical work done by Zhang et al.~\cite{Zhang2015} and experiments by Huebner et al.~\cite{Huebner2018} as well as B\"ohnert et al.~\cite{Boehnert2017b} barrier thermal conductivities at least one order of magnitude below the bulk value or even lower can be expected. Additional optical experiments based on time-domain thermoreflectance~\cite{Cahill2004} could shine light on these buried parts of TMS research. 

The TMS in double-barriers~\cite{Jia2016,Daqiq2018} and the role of magnon transport in the TMS~\cite{Flebus2017} have been discussed recently from the theoretical side. However, experiments following these ideas have not yet been conducted. Finally, the community should move forward towards applications and explore devices that combine the TMS with other spintronic or (magneto)thermoelectric effects to integrate MTJs in nano electronics while recovering unused heat in those nanostructure. Nanostructured sensing devices for three-dimensional thermal gradients~\cite{Martens2018} and the use of thermal gradients in STM devices \cite{Friesen2018b} are only the first realizations of new TMS applications.

\section{Acknowledgments}
First of all we would like to thank all the Master and PhD students as well as postdocs that have contributed to the substantial output of TMS results in our groups in the recent years. These people are namely Alexander Boehnke, Christian Franz, Xiukun Hu, Torsten Huebner, Johannes Christian Leutenantsmeyer, Niklas Liebing, Ulrike Martens, Niklas Roschewsky, Santiago Serrano-Guisan, Marvin von der Ehe (former Walter), Vladyslav Zbarsky, and Jia Zhang. We further kindly acknowledge Karsten Rott for his large effort and support. Most of the TMS results would not be available without his expert knowledge and experience of electron beam lithography and MTJ fabrication. The main part of the work summarized in this review was funded by the priority program SPP 1538 'Spin Caloric Transport' (SpinCaT) of the German Research Foundation (DFG) within the grants KU~3271/1-1, TH~1399/4-2, SCHU~2250/6-2, RE~1052/24-2, HE~5922/4-2, and MU~1780/8-2. Additional financial support by EMRP JRP EXL04 SpinCal is acknowledged. The EMRP is jointly funded by the EMRP participating countries within EURAMET and the EU. We finally thank Stefan Krause and Weiwei Lin for additional input during the revision process of this review article.

\end{document}